# Good Abundances from Bad Spectra: I. Techniques

J. Bryn Jones,[1,2] Gerard Gilmore,[2] and Rosemary F. G. Wyse[2,3,4]

[1] *Department of Physics and Astronomy, University of Wales College of Cardiff, P. O. Box 913, Cardiff, CF2 3YB, Wales*

[2] *Institute of Astronomy, University of Cambridge, Madingley Road, Cambridge, CB3 0HA, England*

[3] *Department of Physics and Astronomy, The Johns Hopkins University, Baltimore, MD 21218, USA*

[4] *Center for Particle Astrophysics, University of California, Berkeley, CA 94720, USA*

21 July 1995

## ABSTRACT

Stellar spectra derived from multiple-object fibre-fed spectroscopic radial-velocity surveys, of the type feasible with, among other examples, AUTOFIB, 2dF, HYDRA, NESSIE, and the Sloan survey, differ significantly from those traditionally used for determination of stellar abundances. The spectra tend to be of moderate resolution (around 1 Å) and signal-to-noise ratio (around 10-20 per resolution element), and cannot usually have reliable continuum shapes determined over wavelength ranges in excess of a few tens of Ångströms. Nonetheless, with care and a calibration of stellar effective temperature from photometry, independent of the spectroscopy, reliable iron abundances can be derived.

We have developed techniques to extract true iron abundances and surface gravities from low signal-to-noise ratio, intermediate resolution spectra of G-type stars in the 4000–5000 Å wavelength region. Spectroscopic indices sensitive to iron abundance and gravity are defined from a set of narrow (few – several Å wide) wavelength intervals. The indices are calibrated theoretically using synthetic spectra. Given adequate data and a photometrically determined effective temperature, one can derive estimates of the stellar iron abundance and surface gravity. We have also defined a single abundance indicator for the analysis of very low-signal spectra; with the further assumption of a value for the stellar surface gravity, this is able to provide useful iron abundance information from spectra having signal-to-noise ratios as low as 10 (1 Å elements).

The theoretical basis and calibration using synthetic spectra are described in this paper. The empirical calibration of these techniques by application to observational data is described in a separate paper (Jones, Wyse and Gilmore 1995). The technique provides precise iron abundances, with zero-point correct to $\sim$ 0.1 dex, and is reliable, with typical uncertainties being $\lesssim$ 0.2 dex. A derivation of the *in situ* thick disk metallicity distribution using these techniques is presented by Gilmore, Wyse and Jones (1995).

# 1  INTRODUCTION

The derivation of chemical abundances for late-type stars is of fundamental importance in studies of stellar astrophysics, Galactic structure and galactic evolution. Abundances of even modest quality provide invaluable data relating to the distribution of stellar populations in the Galaxy and to chemical evolution. The ability to derive useful chemical abundances for faint stars enables the large-scale variations in chemical composition through the history of the Galaxy to be investigated outside the solar neighbourhood.

The strengths of absorption lines in the atmosphere of a star depend on the effective temperature, the stellar surface gravity, and the abundances of different chemical elements. In principle, all these parameters may be measured, given high enough quality data. Indeed, detailed analyses of high signal-to-noise ratio, high-resolution spectra using spectral-synthesis computations and model atmospheres remain the foundation for the determination of accurate stellar chemical compositions. However, while these techniques provide precise abundances for many different elements, the need for high-resolution data demands long integration times on large telescopes for all but the brightest stars. In addition, the necessary slit spectrographs are normally capable of observing only single stars at a time.

Multi-object optical fibre-coupled spectrographs are increasingly being used for radial-velocity surveys of faint objects. Such surveys generate very many spectra, each of which is without reliable continuum flux calibration – due to the difficulty of atmospheric dispersion correction with fibres – and of lower resolution and signal-to-noise ratio than are typically used for stellar abundance determinations. However, full exploitation of such data should include estimations of the stellar metallicities. A technique to achieve this is presented here.

Lower signal-to-noise ratio data require some compromises. For example, it is potentially dangerous to attempt to solve for both temperature and metallicity for late-type stars, since errors in these two parameters are highly correlated, due to the strong sensitivity of metallic line absorption to both parameters. The effective temperature of a star may be estimated from non-spectroscopic data, such as V−I colours, leaving gravity and elemental abundances as the determining parameters of the strengths of absorption features. Defining absorption indices of different sensitivities to abundance and gravity allows a solution for the two parameters to be obtained. This is the philosophy behind the approach advocated here. The problem may be further simplified if one can estimate the surface gravity of the star, for example from an expectation of the evolutionary state, reducing the analysis to a study of the relationship between line strength and elemental abundance. This allows poorer quality data to provide chemical abundance estimates.

We have derived a set of new iron abundance determination techniques, based on abundance indices, that is presented here. These indices provide a true iron abundance and are optimised for the study of low signal-to-noise (around 10 in 1Å pixels), intermediate-resolution (around 1Å) spectra in the 4000Å to 5000Å wavelength range, but can be used successfully for data of lower resolutions. The effects of noise in the intensity data may be mitigated by use of a large number of such indices. Indices sensitive to both iron abundance and surface gravity have been selected. Full details of the development of the techniques may be found in Jones (1991). Provided that a spectrum is not too noisy, it is possible to solve for both iron abundance and surface gravity, given a photometrically determined effective temperature. The theoretical calibration of these techniques through spectral synthesis computations is presented in this paper.

The stellar spectral synthesis computations are described in Section 2. Atomic and stellar parameters were carefully selected to reduce the possibilities of calibration errors. The selection criteria for the absorption line indices are discussed in Section 3. The final set of 16 abundance and gravity indices are defined, based on 80 flux bands. Section 4 presents the calibration and testing of these indices using synthetic spectra, first noise-free, then with noise added to simulate real data. The derivation of abundance information from particularly noisy spectra through the assumption of a surface gravity is considered. Section 5 presents a discussion of the errors which are likely to affect an abundance measurement.

## 2  THE COMPUTATION OF SYNTHETIC SPECTRA

### 2.1  The spectral synthesis methods

The determination of the basic atmospheric physical parameters of stars in these techniques is based on a comparison of an observed spectrum with information obtained from theoretical simulations of spectra. These theoretical data required the computation of an extensive grid of synthetic spectra covering the range of stellar parameters of interest. The strengths of observed absorption features are compared with the strengths of the same features in the synthetic data to derive relevant atmospheric parameters. The theoretical spectra were computed through the solution of the equation of radiative transfer at a grid of wavelengths using an upgraded version of an atomic line LTE spectral synthesis program written by

P. M. Williams in the early 1970's. A standard LTE line formation theory was used (*e.g.* Mihalas, 1970, Gray, 1976).

## 2.2 Atomic and stellar data

The metallicity determination methods were developed for the analysis of spectra of mainly moderately metal-poor G dwarfs; thus a temperature-scaled Holweger–Müller (1974) solar model atmosphere formed the basis of the technique. The validity of the model has been confirmed by studies of the solar atmosphere (*e.g.* Grevesse, 1984) and, after scaling, of the atmospheres of G-type dwarfs. This scaled model provided the basic relation between temperature and the optical depth at a wavelength of 5000Å through the atmosphere.

Functional representations of atomic partition functions were taken from the extensive compilation of Irwin (1981). The classical approximation for the natural broadening parameter was used. Stark electronic and ionic broadening was neglected as all lines included in the synthetic spectra are caused by heavy elements, for which this effect is not important in stars of the temperatures of interest.

Solar photospheric elemental abundances were taken from the review of Grevesse (1984), including an iron abundance relative to hydrogen of $N(Fe)/N(H) = 4.7 \times 10^{-5}$ and a helium abundance of $N(He)/N(H) = 0.10$. For simplicity, these solar relative abundances were adopted for all stellar models, scaling the elemental abundances in step with metallicity. However, rather tight trends in element abundance ratios are known for Galactic stars (*e.g.* Wheeler *et al.* 1989, Edvardsson *et al.* 1993), which will affect the abundance scale of the present analysis through the contributions of contaminating species to regions of iron line absorption, and through the general contribution of heavy elements to the opacity in the atmosphere. These effects are removed through the option of an empirical recalibration of the final iron abundance scale, rather than by attempting direct modelling.

A single, depth-independent microturbulence parameter was adopted for all stars. In the case of the disc-centre solar spectrum the value was taken to be $\xi_{micro} = 0.85$ kms$^{-1}$, following Blackwell *et al.* (1984) (from Fe I lines) and Blackwell *et al.* (1987) (from Ti I lines). For the solar flux (integrated disc) spectrum a microturbulence of $\xi_{micro} = 1.18$ kms$^{-1}$ was adopted (Blackwell *et al.*, 1987). A microturbulent parameter of $\xi_{micro} = 1.5$ kms$^{-1}$ was assumed for all stars other than the Sun. This value is typical of recent analyses of solar-type dwarfs (*e.g.* Magain, 1989, Hartmann and Gehren, 1988, Gratton and Sneden, 1987), and is consistent with the narrow-band spectroscopic index study of Nissen (1981). This value was used for stars of all surface gravities; abundance errors caused by an inappropriate choice of the microturbulence parameter are considered in Section 5.4.

## 2.3   Collisional broadening of spectral lines

Our spectroscopic analysis methods discussed here employ the pressure sensitivity of selected strong Fe I absorption lines to derive surface gravity information. Since these lines lie on the damping part of the curve of growth, collisional broadening mechanisms are of fundamental importance in determining their line strengths. As such, it is essential that the collisional broadening processes be represented accurately during the spectral synthesis computations. Even in metal-poor stars where a particular spectral line may lie on the linear or saturated parts of the curve of growth, appreciable errors may be introduced through the oscillator strength data if performing a differential analysis when the equivalent line suffers from the effects of damping in the solar spectrum (Magain, 1984). Following conventional practice we computed the collisional damping parameter $\Gamma_{coll}$ using the Unsöld approximation (*e.g.* Mihalas, 1970, Böhm, 1960, Blackwell & Collins, 1972), correcting this approximate value, $\Gamma_6$, with an empirical enhancement factor $E$, to give $\Gamma_{coll} = E \, \Gamma_6$. Although the Unsöld approximation predicts a dependence on temperature of $\Gamma_6 \propto T^{0.3}$, laboratory studies have shown that a slightly stronger temperature sensitivity is more realistic. Following the recommendation of Lwin *et al.* (1977) for broadening of metal atom lines by light perturbers, a $T^{0.4}$ temperature dependence has been adopted, using a normalisation of the collisional damping parameter per unit hydrogen atom number density at a temperature of 5000 K (*e.g.* O'Neill and Smith, 1981).

We selected enhancement factors carefully, based on a critical review of published data. Although the available data were reviewed by Gurtovenko and Kondrashova (1980), the usefulness of many of these results is limited by the low accuracy of the oscillator strengths and model atmospheres used. Among the more careful studies of solar Fe I collisional damping, Simmons and Blackwell (1982) used the Holweger-Müller solar model atmosphere and accurate oscillator strengths to fit computed line profiles to the observed solar profiles of 24 Fe I lines, while Gratton and Sneden (1988) considered 11 lines. Seven Fe I lines common to both studies allow the consistency of the two enhancement scales to be assessed, showing that the Gratton and Sneden values are on average 5% larger, with a scatter is 10%. The empirical estimates of damping of Bell, Edvardsson and Gustafsson (1985) for two Fe I lines in the solar spectrum are close to the results for the same two lines of Simmons and Blackwell and of Gratton and Sneden.

We chose here to express the enhancement factor as a function of the excitation potential, rather than of multiplet number, to minimise the number of parameters required to describe each atomic transition. Figure 1 shows the available data from the literature (sources identified in the Figure caption) in a plot of enhancement factor against lower excitation potential, $\chi_{exl}$. A simple mean relation can be found for the higher quality data based on the accurate laboratory oscillator strengths, but the behaviour for $\chi_{exl} > 3.0$ eV is less clear. For convenience, the data will be represented by a straight line passing through the points with $E = 1.0$ for transitions from the ground level (consistent with the mean relation of Simmons and Blackwell, 1982), and $E = 1.4$ at $\chi_{exl} = 3.0$ eV. For larger excitations, a constant enhancement of $E = 1.4$ will be adopted. The enhancement for Fe I is therefore given as

$$\begin{aligned} E_{FeI} &= 1.0 + 1.33\,(\text{eV})^{-1}\,\chi_{exl} \quad \text{for } 0.0\,\text{eV} \leq \chi_{exl} \leq 3.0\,\text{eV}\ ,\\ &= 1.4 \quad \text{for } \chi_{exl} > 3.0\,\text{eV}\ . \end{aligned} \quad (1)$$

This representation is shown in Figure 1. It is slightly less steep than the mean interpolated relation adopted by Simmons and Blackwell.

It should be noted that most of the Gurtovenko, Fedorchenko and Kondrashova lines arise from transitions where the lower term has odd parity, in contrast to the data of Simmons and Blackwell and of Gratton and Sneden. However, the mean enhancement of their (odd parity) data is comparable to that of the even parity values of Simmons and Blackwell and of Gratton and Sneden at lower excitations. The mean relations given above will be adopted for all lines.

The accurate representation of collisional broadening for absorbing species other than Fe I is of lesser importance for the present analysis, since the wavelength regions used to define flux bands of spectroscopic indices are chosen specifically to exclude strong lines of other species, as discussed below in Section 3. Despite these less stringent requirements, an attempt was made to review published damping parameter data for these other species. As in the case of Fe I collisional line broadening, some empirical determinations of damping parameters are available based on fitting line profiles in the solar spectrum. Some laboratory measurements, usually of the perturbing effects of helium atoms, have provided estimates of line broadening by collisions with hydrogen atoms.

The data are limited in both quantity and quality, and so it is not possible to attempt to represent the variations of enhancements with atomic parameters; they are assumed to be constant for each species. Typical mean enhancement factors are presented in Table 1. Default values of $E = 1.5$ are adopted when no useful empirical estimates are available. The use of mean enhancements for each species is unlikely to cause significant errors in this study, where only the combined absorption effects of large numbers of lines are of interest.

## 2.4 Testing the synthesis methods

The reliability of the spectral synthesis computations was tested by comparing synthesis results both with the observed solar spectrum and with the published stellar analysis results of other authors carried out using their synthesis software. Consistency was shown in all cases.

Several unblended Fe I, Cr I and Ti I lines with accurate oscillator strengths in the solar disc-centre spectrum were synthesised and compared with the observed high resolution solar spectrum from the Delbouille *et al.* (1973) atlas. The synthetic profiles were in very good agreement with the observations, with the sole exception of the Cr I line at 4942.48 Å . Assuming that the oscillator strength value is correct, one possible explanation for the discrepancy between the synthetic spectrum and the observations is that the Cr I line is blended with the neighbouring Mn I line. This is discussed more fully by Jones (1991).

The solar iron abundance analysis of Blackwell, Booth and Petford (1984) and the titanium abundance analysis of Blackwell, Booth, Menon and Petford (1987) were repeated, adopting the published equivalent widths. The results of the present synthesis methods were in very good agreement with the published data.

The published analyses by Magain (1984) for two very metal-poor stars, one [Fe/H] = −2.3 dwarf (HD 19445) and one [Fe/H] = −3.0 subgiant star (HD 140283) provides a check of the method for stars whose atmospheric parameters differ substantially from those of the Sun. Whereas Magain used LTE model atmospheres by Gustafsson which were selected for appropriate stellar parameters, in this reanalysis the Holweger-Müller atmosphere was scaled within the synthesis program to the Magain stellar parameters. This provided an assessment of the error in the abundance scale arising from the use of a scaled solar model.

Atomic and line data from his quoted references were used, including the Blackwell group oscillator strengths for all but the 4528.63 Å line for which the value was taken from Magain (1985). Magain's (1984) Fe I equivalent widths were used, except for the 4494.57 Å line in HD 19445 for which the revised value of Magain (1985) was adopted. The mean difference between the [Fe/H] results of the synthesis program and those of Magain for HD 19445 was +0.15 ($\sigma$ = 0.02), while for the 5420 K reanalysis of HD 140283 the mean difference in [Fe/H] between the program results and those of Magain was +0.19 ($\sigma$ = 0.03). When it is considered how extreme the atmospheric parameters of these stars are when compared to those of the stars of interest to this work, the agreement is satisfactory.

Thus in all cases the synthesis program was shown to work more than adequately for the purposes of this study.

## 2.5 Spectral line data

Computation of the profile of a spectral line requires a minimum number of basic parameters describing the line. It is possible to limit these to only four items of information: the identification of the atomic or ionic species responsible for the absorption, the central wavelength of the line, the lower excitation potential of the electronic transition, and the intrinsic strength of the absorption process. We require the appropriate data for many hundreds of spectral lines, which we adopt from Moore, Minnaert and Houtgast (1966).

The data for all the atomic and ionic lines listed by Moore *et al.* in the wavelength ranges of interest were used as input to the synthesis program. The wavelength regions used to define metallicity indicators were chosen for the virtual absence of molecular absorption; molecular lines are not synthesised and so those data were not included. Lines which are listed as being present in sunspot spectra only were also ignored. The numbers of the lines used in each region are listed in Table 2.

Some blended lines are given only a single, integrated equivalent width in the Moore *et al.* list, but usually with individual estimates of their reduced widths (defined as the ratio of the equivalent width to wavelength). Estimates of their individual equivalent widths have been calculated assuming, naïvely, that the relative contributions of each line to the integrated equivalent width are in direct proportion to their reduced widths. When no indications are given of the relative line strengths of two or more lines in a blend, it has been assumed that they contribute equally to the integrated equivalent width. If one or more of the components of a blend is listed as being dominant then only these components are considered.

One possible difficulty which may arise from using the Moore *et al.* data is that caused by wrongly identified spectral lines. However, the techniques developed here rely on the integrated absorption from very large numbers of lines and we assume that the effects of random errors largely cancel themselves out.

A proportion of the lines in the Moore *et al.* list have incomplete data, lacking excitation potentials and even identifications. These are almost without exception weak lines, often with solar equivalent widths of only a few mÅ. They are, however, relatively common; for example, of the 220 lines between 4950Å and 5000Å, 99 lack full data. Their contribution to the total absorption in solar-type stars will be generally small, but it is desirable to attempt to account for it. The most common absorbing species in the visible spectra of solar-type stars is neutral iron. It is likely that many of the unidentified lines will be due to Fe I transitions (at least in the wavelength regions considered here which are relatively free from molecular contamination) and therefore all such lines will be attributed to Fe I for the purposes of this work. The mean value of the lower excitation potential of all Fe I lines between 4900Å and 5000Å having complete line data in the Moore *et al.* list is 3.7 eV; following Nissen (1970a), we adopt a value of 4.0 eV.

The behaviour of spectral lines is strongly dependent on excitation potential; it is important to confirm that attributing unidentified lines to 4.0 eV Fe I transitions does not cause systematic errors in accounting for line absorption in stars significantly different from the Sun. We therefore investigated the effects of the assumed identification of these weak lines through a detailed investigation of the 4938Å – 4946Å wavelength region, for which Moore *et al.* list a total of 41 absorption lines of which 20 are unidentified (and 21 of the 41 are without excitation potentials). A solar disc-centre spectrum was computed, assuming that unidentified lines were caused by 4.0 eV Fe I transitions, calculating oscillator strengths by iteratively adjusting the values for each line until adequate fits were obtained to the profiles of the Delbouille *et al.* (1973) solar atlas. Oscillator strength data from the Blackwell group were available for two lines. The fit is generally very good, with the exception of the 4942.48 Å Cr I line, which, as discussed above, is a perennial problem.

This set of atomic data was then used to synthesise the spectrum of Arcturus using the atmospheric parameters of Mäckle *et al.* (1975). This is compared with smoothed profiles from the Griffin (1968) atlas in Figure 2. The star has generally stronger metallic line ab-sorption than that of the Sun, easing the study of lines which are weak in the solar spectrum and allowing the consequences of an appreciable change in atmospheric parameters to be determined. The cores of strong lines are too deep in the synthetic spectrum; this may be partially due to the fact that for convenience, gaussian instrumental and macroturbulence profiles were used. There are no appreciable systematic errors in the profiles of the unidentified lines, although the synthetic lines are sometimes too strong and at other times too weak. This confirms that the practice of attributing unidentified lines to 4.0 eV Fe I transitions will not introduce significant systematic errors.

## 2.6 The Determination of Oscillator Strengths

Oscillator strengths were computed from the solar spectrum for a majority of lines. As such these techniques essentially perform a differential analysis. Solar disc centre equivalent widths were taken from the Moore, Minnaert and Houtgast (1966) compilation, recalibrating their equivalent width scale to account for the possibility of systematic errors. Selected accurate weighted oscillator strength data, $gf$, were taken for the literature for some stronger lines, but these represented only a small minority of the total number of absorption lines considered.

The recalibration of the Moore, Minnaert and Houtgast equivalent width scale was effected by comparison with accurate measurements of a number of absorption lines in the solar disc-centre spectrum from the Blackwell group (Fe I: Blackwell, Booth and Petford, 1984, and, for strong lines, from Blackwell and Shallis, 1979; Ti I and Cr I: Blackwell, Booth, Menon and Petford, 1987). The Moore *et al.* equivalent widths, $W_{\lambda\,MMH}$, are plotted against the accurate data, $W_{\lambda\,acc}$, in Figure 3. The relation $W_{\lambda\,MMH} = W_{\lambda\,acc}$ is shown for comparison. A quadratic relation was fitted by least squares, giving the very strong lines increased weights to prevent the larger numbers of weaker lines from disproportionately influencing the relation at large equivalent widths. The resulting re-calibration was:

$$W_{\lambda\,MMH} = 3.747 \times 10^{-4}\,(\text{m\AA})^{-1}\,W^2_{\lambda\,acc} +$$
$$0.8949\,W_{\lambda\,acc}$$
$$\text{for}\ \ 0\ \leq\ W_{\lambda\,MMH}\ \leq\ 280\ \text{m\AA}\ ,$$

$$\text{and}\ \ W_{\lambda\,MMH}\ =\ W_{\lambda\,acc}\ \ \text{for}\ \ W_{\lambda\,MMH}\ >\ 280\ \text{m\AA}\ .$$

where we assume an identity between the scales beyond 280mÅ. This relation is also plotted in the figure.

The computed *gf*-values were replaced by accurate laboratory data when these were available, as discussed below. On a few occasions, a laboratory measurement was available for a line in a blend for which only an integrated equivalent width is given by Moore *et al.*. In these cases, the accurate value was used to synthesise the solar disc-centre line strength. This was then subtracted from the integrated value to give the strength of the remaining line or lines, ignoring saturation effects. New *gf*-values were calculated for these remaining lines.

More accurate equivalent width measurements are available in the literature for some of the spectral lines of interest (e.g. Gurtovenko and Kostik, 1982, list Fe I solar disc-centre equivalent widths). When available these were used in preference to the recalibrated Moore *et al.* data to compute oscillator strengths. This procedure was adopted instead of using the published Gurtovenko and Kostik *gf*-values, in order to avoid any systematic differences which might otherwise arise. Fe II line data were taken from Blackwell, Shallis and Simmons (1980) if available, again computing *gf*-values from solar line strengths. We did not use published solar data that were not available as equivalent widths. Laboratory oscillator strength results were used instead of solar estimates for lines for which *gf*-value data had been published by the Blackwell group for Fe I, Ti I, Ti II, Cr I and Mn I transitions up to the end of 1988, and by Smith and Raggett (1981) for Ca I lines.

## 2.7    Computation of synthetic spectra

Synthetic spectra with a range of gravities, metallicities and effective temperatures were used to select the index flux bands, in addition to their use in calibrating the analysis techniques. Stellar parameters were chosen to form a rectangular grid in parameter space, as listed in Table 3. The element ratios are fixed at the solar values. Five different temperatures, four gravities, five metallicities and one microturbulence ($\xi_{micro}$ = 1.5 kms$^{-1}$) were used; the basic grid of stellar models contains 100 points. Some additional spectra were computed having a microturbulence of $\xi_{micro}$ = 1.0 kms$^{-1}$; these were used in the error analysis of Section 5. The wavelength regions chosen are described in Section 3.4. The spectra were computed in individual sections each about 100Å wide.

All the synthetic stellar spectra were produced in the form of integrated-disc fluxes, normalised to a continuum level of unity, at wavelengths on a regular 0.01 Å interval grid. The effects of macroturbulent velocities, rotation or instrumental broadening were not added at this stage. Solar disc-centre spectra were computed initially, allowing a check for gross errors in the synthesis techniques to be made by comparing the line profiles with the Delbouille, Roland and Neven (1973) atlas.

## 3 THE SELECTION OF INDICES SENSITIVE TO STELLAR PARAMETERS

### 3.1 The methods used to define indices

Spectroscopic indices measure stellar physical parameters by comparing features in the spectrum which are sensitive to the parameters of interest with other regions which are either less sensitive or show an opposite dependence. It is normal to use an index defined to be the ratio of the flux in a region of the spectrum which contains appropriate features, to the flux in a comparison band. This may be extended to use more than two wavelength regions, which has the advantage of combining greater flexibility with the use of a larger total width of spectrum in the index, reducing errors arising from random noise in the data. Thus we define an index as the ratio of the sum of the fluxes in a number of different bands which are sensitive to the desired quantities, to the sum of the fluxes in a number of comparison bands. Wavelength regions which measure spectral features will be referred to as absorption bands.

The importance of using as large a total wavelength range as possible must be emphasised. This is particularly true for the analysis of the spectra in this work where the dominant source of error is the random noise in the signal intensities. The relative error in a flux measurement is reduced as the band width is increased, resulting in a decrease in the relative error in the index value. However, frequently, the features in the spectrum are narrow and increasing the width of the flux bands will not provide any advantages. Instead, an increase in the width merely adds insensitive regions to the flux band and therefore decreases the sensitivity of the index. The selection of flux bands often requires a compromise between these two effects.

The techniques adopted for this work involve the calculation of fluxes by the integration of intensity over wavelength between two wavelength limits. The flux bands therefore possess sharp edges, modified only by the effects of instrumental resolution, in contrast to the smooth response curves which are encountered in photometric studies. Consequently, the bands used here are the equivalent of what would be rectangular transmission curves in photometric work. Fluxes from rectangular bands are more sensitive to wavelength scale errors than those from smooth-edged bands.

The variation in the continuum level between flux bands can introduce systematic errors into the calibration and interpretation of index data. As described in more detail below, the continua of the spectra are re-normalised to a constant level. Optimally, one chooses the flux bands of an index so that they lie within a relatively small wavelength range of each other; for the present application, this is a few tens of Ångström units.

For the present study, we assume linear detectors, so that available intensities are real intensities. The flux $F$ in a wavelength band is therefore defined to be the integrated detected signal within the band and is given by

$$F = \int_{\lambda_1}^{\lambda_2} I_\lambda(\lambda) \, d\lambda \quad , \tag{2}$$

where $I_\lambda(\lambda)$ is the signal intensity at a wavelength $\lambda$, while $\lambda_1$ and $\lambda_2$ are respectively the short and long wavelength limits of the band.

It is convenient to normalise fluxes to give the *residual flux*, defined to be the ratio of the flux in a band to the value which would be obtained in the absence of line absorption. For a constant continuum level $I_{\lambda c}$, the residual flux in a band is given by

$$f = \frac{F}{I_{\lambda c} \, \Delta\lambda} \quad , \tag{3}$$

where $\Delta\lambda \equiv (\lambda_2 - \lambda_1)$ is the width of the band in wavelength units. Therefore, $f = 1$ in the absence of absorption, and $f < 1$ in the presence of absorption in the band.

The flux described by Equation 2 represents the case where the intensity is available as a continuous function. Observed spectra, however, are most often expressed as the fluxes in a large number of narrow wavelength bins. The total flux in a band is therefore the sum of the individual fluxes in the pixels within its wavelength range. Neglecting problems caused by wavelength limits falling within pixels, the total flux is given by

$$F = \sum_{i=1}^{N_p} F_{p\,i} \quad , \tag{4}$$

where $F_{p\,i}$ is the flux in the $i^{th}$ pixel in the flux band, and $N_p$ is the total number of pixels in the band.

An observed spectrum will suffer from random noise. The flux in each pixel will suffer from an error described by the signal-to-noise ratio of the data. The error is therefore given by

$$\frac{\text{error in } F_{p\,i}}{F_{p\,i}} = \frac{1}{R_{S/N\,p}} \quad , \tag{5}$$

where $R_{S/N\,p}$ is the signal-to-noise ratio per pixel, which is assumed to be constant across the spectrum. When the absorption is small, the pixel flux may be approximated by the value, $F_{p\,c}$, which would be obtained in the absence of absorption (the continuum level pixel flux). Therefore

$$\begin{aligned}\text{error in } F_{p\,i} &= \frac{F_{p\,c}}{R_{S/N\,p}} \\ &= \frac{I_{\lambda\,c}\,w_p}{R_{S/N\,p}} \quad ,\end{aligned} \tag{6}$$

where $w_p$ is the width of the pixel. The error in the total flux can be expressed in terms of the error in each pixel using a simple error analysis on Equation 4. This gives

$$\text{error in } F = \sqrt{\sum_{i=1}^{N_p} (\text{error in } F_{p\,i})^2} = \sqrt{\sum_{i=1}^{N_p} \left(\frac{I_{\lambda\,c}\,w_p}{R_{S/N\,p}}\right)^2} \quad , \tag{7}$$

on substituting the pixel error from Equation 6. For a constant continuum level, pixel width and signal-to-noise ratio,

$$\text{error in } F = \frac{I_{\lambda\,c}\,w_p}{R_{S/N\,p}} \sqrt{N_p} \quad . \tag{8}$$

The number of pixels in the flux band, $N_p$, is related to the band and pixel widths by $\Delta\lambda = N_p\,w_p$. On substituting for $N_p$ into (8), the final result for the error in the total flux is

$$\text{error in } F = \frac{I_{\lambda\,c}\sqrt{w_p\,\Delta\lambda}}{R_{S/N\,p}} \quad . \tag{9}$$

Similarly, it is possible to express the error in the residual flux in a band in terms of the signal-to-noise ratio and the width of the band. Using Equation 3, the error in the residual flux is given by

$$\text{error in } f = \frac{\text{error in } F}{I_{\lambda\,c}\,\Delta\lambda} \quad .$$

Substituting from 9, the final result obtained is

$$\text{error in } f = \frac{1}{R_{S/N\,p}} \sqrt{\frac{w_p}{\Delta\lambda}} \quad . \tag{10}$$

An index may be represented in terms of the ratio $R$ of the sum of the fluxes in absorption bands to the sum of those in comparison bands. The ratio is

$$R = \frac{\text{total flux in absorption bands}}{\text{total flux in comparison bands}} = \frac{\sum_{i=1}^{N_A} F_{A\,i}}{\sum_{i=1}^{N_C} F_{C\,i}} \quad , \tag{11}$$

where $F_{A\,i}$ is the flux in the $i^{th}$ absorption band, $F_{C\,i}$ is the flux in the $i^{th}$ comparison band,

$N_A$ is the total number of absorption flux bands, and $N_C$ is the total number of comparison flux bands used to define the index. This may, alternatively, be expressed in terms of the residual fluxes by substituting for the fluxes from Equation 3 .

$$R \;=\; \frac{\sum_{i=1}^{N_A} f_{A\,i}\, \Delta\lambda_{A\,i}}{\sum_{i=1}^{N_C} f_{C\,i}\, \Delta\lambda_{C\,i}} \quad,$$

where $f_{A\,i}$ is the residual flux in the $i^{th}$ absorption band, $f_{C\,i}$ is the residual flux in the $i^{th}$ comparison band, $\Delta\lambda_{A\,i}$ is the width of the $i^{th}$ absorption band in wavelength units, and $\Delta\lambda_{C\,i}$ is the width of the $i^{th}$ comparison band in wavelength units. For convenience, the index will be set to be proportional to the ratio $R$ and the constant of proportionality chosen so that the index has a value of unity in the absence of absorption. Without absorption, the residual fluxes are $f_{A\,i} = f_{C\,i} = 1$ for all $i$. The index is therefore defined to be

$$I \;\equiv\; \frac{\left(\sum_{i=1}^{N_C} \Delta\lambda_{C\,i}\right)\, \sum_{i=1}^{N_A} f_{A\,i}\, \Delta\lambda_{A\,i}}{\left(\sum_{i=1}^{N_A} \Delta\lambda_{A\,i}\right)\, \sum_{i=1}^{N_C} f_{C\,i}\, \Delta\lambda_{C\,i}} \quad. \tag{12}$$

In terms of the true fluxes this is

$$I \;=\; \frac{\left(\sum_{i=1}^{N_C} \Delta\lambda_{C\,i}\right)\, \sum_{i=1}^{N_A} F_{A\,i}}{\left(\sum_{i=1}^{N_A} \Delta\lambda_{A\,i}\right)\, \sum_{i=1}^{N_C} F_{C\,i}} \quad. \tag{13}$$

The error in the index value caused by errors in the fluxes can be estimated by means of the propagation of errors. In the case where the absorption is small, $I \simeq 1$, $f_{A\,i} \simeq 1$ and $f_{C\,i} \simeq 1$ for all $i$. On using the approximation $I = 1$, $f_{A\,i} = 1$ and $f_{C\,i} = 1$ for all $i$, the index error becomes

$$\text{error in } I \;=\; \sqrt{\frac{\sum_{i=1}^{N_A} \Delta\lambda_{A\,i}^{\,2}\, (\text{error in } f_{A\,i})^2}{\left(\sum_{i=1}^{N_A} \Delta\lambda_{A\,i}\right)^2} + \frac{\sum_{i=1}^{N_C} \Delta\lambda_{C\,i}^{\,2}\, (\text{error in } f_{C\,i})^2}{\left(\sum_{i=1}^{N_C} \Delta\lambda_{C\,i}\right)^2}} \quad,$$

and the errors in the residual fluxes become (Equation 10)

$$\text{error in } f_{A\,i} \;=\; \frac{1}{R_{S/N\,p}}\, \sqrt{\frac{w_p}{\Delta\lambda_{A\,i}}} \quad;\qquad \text{error in } f_{C\,i} \;=\; \frac{1}{R_{S/N\,p}}\, \sqrt{\frac{w_p}{\Delta\lambda_{C\,i}}} \quad.$$

This is a relatively good approximation for the spectra considered here. The index error now becomes

$$\text{error in } I \;=\; \frac{\sqrt{w_p}}{R_{S/N\,p}}\, \sqrt{\frac{1}{\sum_{i=1}^{N_A} \Delta\lambda_{A\,i}} + \frac{1}{\sum_{i=1}^{N_C} \Delta\lambda_{C\,i}}} \quad. \tag{14}$$

It is these definitions which will be used throughout. That of Equation 12 is appropriate for the calculation of the index values from residual fluxes which have been derived from synthetic spectra. The observed index may be calculated from the observed fluxes after intensity scale normalisation using Equation 13. The error in an index value is given by Equation 14 when the degree of absorption is small; this result will be used below to define alternative sets of indicators from those discussed here. Equation 14 clearly illustrates the need to use as great a total wavelength range as possible, when defining both absorption and comparison flux bands, in order to reduce the error in an index value, as described earlier. We note here that this error, together with all other used in this paper, corresponds to a standard error where the parent distribution is Gaussian. That is, the quoted error is the half-width at $exp^{-0.5}$ of the maximum of an assumed normal distribution.

Having established the manner in which indices will be calculated from appropriate regions of a stellar spectrum, it is now necessary to select the regions of interest. The remainder of this Section is devoted to the choice of flux bands.

### 3.2 Criteria for the selection of ideal indices

An assumption that is often made during the selection of flux bands of indices is that the regions of the spectrum which experience the greatest absorption, also possess the greatest sensitivities to elemental abundances. Absorption bands are therefore normally chosen to contain the strongest absorption features. Nearby regions of the spectrum serve as comparison bands. This, however, may not necessarily lead to the selection of indices which either have a maximum sensitivity to iron abundance or which allow errors to be minimised. These points are considered in turn below.

The sensitivity of a spectral line to abundance is dependent on its position on the curve of growth. There would be distinct advantages in selecting abundance indices in which the absorption band contained a considerable number of lines on the linear part of the curve of growth, in contrast to a smaller number of stronger lines contributing a similar amount of total absorption. The sensitivity to abundance is maximised for the former case. Equally significantly, the errors in the theoretical calibration of indices caused by inadequate knowledge of microturbulence and damping parameters are reduced. An ideal iron abundance index therefore should have an absorption band containing as many Fe I lines on the linear part of the curve as possible, with few stronger Fe I lines and a minimal degree of contamination by other species. The comparison band should be relatively free of absorption, and again suffer little contamination from species other than Fe I. Nissen (1970b) identified an index around 4800 Å which satisfies these requirements for late F-type stars.

If a differential analysis is being performed, where oscillator strength data are calculated from the spectrum of a standard star, the errors in the $gf$-values will be reduced if the lines in the flux bands of an index lie on the linear part of the curve in both the standard star and that being analysed. However, during studies of solar-type metal-poor stars, it often becomes impossible to ensure this. This is a problem which is also encountered during conventional high-resolution analyses of metal-poor objects (as discussed, for example, by Peterson and Carney, 1979).

A complete search for regions of the spectrum that could be used to define indices based on these ideal criteria was made through the Moore, Minnaert and Houtgast (1966) line list over the full 4000 to 5000 Å wavelength range. This contains very many closely-spaced absorption lines; neutral iron lines dominate throughout, although Cr I, Ti I and Ni I are also plentiful and are important contributors to the total absorption. There are some very strong lines present, such as the Fe I lines at 4045 Å and 4384 Å, particularly at shorter wavelengths. Considerable numbers of molecular lines are also present, although these molecular bands are confined to certain wavelength ranges. Index flux bands must be essentially free of molecular absorption. The wavelength regions affected by molecular absorption are listed in Table 4.

Very few wavelength regions were found which satisfied the ideal selection criteria. Rather, there are very large numbers of absorption lines on the saturated or damping parts of the curve of growth in any broad region of the solar spectrum. It becomes difficult to identify wavelength intervals of the length required for flux bands ($\gtrsim$ few Å) which possess considerable numbers of lines on the linear part of the curve-of-growth, without appreciable contamination by stronger lines. It is virtually impossible to find wavelength intervals which also possess enough absorption to allow indices of adequate sensitivity to be defined. It is also comparatively difficult to find intervals which are sufficiently free of absorption that they could be used as comparison regions for these ideal indices. It must be stressed that the analysis of low-signal spectra requires the use of large numbers of indices to reduce the effects of random errors. Consequently, any selection criteria used must allow sufficient numbers of indices to be found.

The spectra of metal-poor G-type stars may, depending on temperature, have absorption by neutral species of iron group elements which is considerably weaker than in the solar spectrum. Flux bands for metal-poor stars should contain spectral lines which are generally stronger in the solar spectrum than those unsaturated lines discussed above. Although this eases the problems encountered in finding indices during a search of the Moore *et al.* list, it is still difficult to identify sufficient numbers of bands which satisfy the ideal index criteria.

It was therefore not possible to identify adequate numbers of flux bands to allow sufficient indices to be defined based on the ideal selection criteria. Less stringent criteria had to be used to choose the bands which were required to establish the metallicity estimation techniques.

### 3.3 Criteria for the selection of practical indices

A first step in identifying practical indexes is to determine which regions of the spectra of the stars of interest show the greatest sensitivity to metallicity, and to surface gravity. For this purpose, synthetic spectra were computed for the 4900Å – 5000Å wavelength range, using the same basic techniques that were discussed in Section 2. Initially the spectrum of a solar metallicity G0 dwarf was synthesised, together with two additional spectra which were identical to it except for either an increased metallicity (greater by +0.3 in [Fe/H]) or a decreased surface gravity (smaller by 0.3 in $\log_{10} g$). The spectra were broadened to a resolution of 1.0Å (by convolution with a 1.0Å full-width at half-maximum gaussian instrumental profile). The sensitivities of each region in the G0 V star spectrum to the changes in metallicity and in gravity could be determined by calculating the difference in residual intensity between the first spectrum and those with altered parameters.

There is a very strong correlation between the metallicity sensitivity and the strength of the absorption in the spectrum, as expected from the discussion above. This implies that abundance indices should be selected by identifying the regions of the spectrum which have the greatest iron line absorption and little contamination by other species to serve as the absorption flux bands, and the regions which have little absorption to serve as the comparison bands.

The sensitivities of different regions of the spectrum to surface gravity vary relatively little. There is some tendency for the stronger absorption regions to have a greater dependence on gravity, although some spectral features, particularly ionic lines, have a dependence in the opposite direction (decreasing rather than increasing in strength with increasing gravity). The most noticeable feature is the existence of a few narrow regions which show very strong sensitivities. These correspond to very strong spectral lines, with the gravity dependence being caused by the sensitivity of the damping wings to gas pressure. The pressure sensitivities of very strong Fe I lines have indeed been proposed as a means to determine stellar gravity (*e.g.* Blackwell and Willis, 1977, and Bell, Edvardsson and Gustafsson, 1985).

Abundance indices defined in regions of the spectrum which contain strong absorption features will have only a slight sensitivity to gravity if the absorption is due to numbers of moderately strong lines (but not the very large numbers of weaker lines as demanded by the ideal selection criteria). If the absorption is produced by a smaller number of very strong lines, even a single line in some cases, there will be a strong sensitivity to gravity. Therefore the possibility exists for the selection of two types of abundance index according to the strengths of the spectral lines present, one type being essentially insensitive to surface gravity, and the other showing a considerable dependence. In addition to these metallicity-sensitive indices, there is a possibility that indices sensitive to gravity alone can be defined, based on a comparison of the strength of ionic lines with absorption by neutral atoms. However, these effects are relatively weak and the numbers of such lines are limited. Indices of this type are therefore likely to suffer from considerable errors during the analysis of noisy spectra.

### 3.4  A list of provisional abundance indices

As discussed in the preceeding section, abundance indices need to be based on regions of strong iron line absorption. Ideally, a list of indices could be produced by identifying regions of strong absorption, then confirming that these are caused by Fe I lines and that the degree of contamination by other species is small. A second detailed search was made through the Moore *et al.* (1966) solar line list. Regions of strong absorption were noted, including those with very strong lines and those with numbers of weaker lines grouped together. Contamination by species other than Fe I was also noted. A list of best flux indices was compiled using the selection criteria discussed in the preceeding section. This allowed a list of provisional abundance indices to be produced, using a total of 29 flux bands. The list is presented in Table 5.

The need to avoid molecular lines excluded large regions of the spectrum from consideration during the choice of indices, particularly the section with wavelengths between 4150Å and 4420Å. The region lying between the H$\beta$ line and the long wavelength limit of the techniques at 5000Å proved particularly fruitful for provisional indices. Although the absorption is not particularly strong in this 4875Å – 5000Å section compared with shorter wavelengths, the spectrum can be broken into useful absorption and comparison bands. It is possible to define some comparatively sensitive indices from these. The 4500Å – 4690Å region possesses a moderate degree of line absorption and provides three provisional indices. A number of species other than Fe I are found to produce lines in this region and ionic lines are more common than in the 4875Å – 5000Å section. However, the division of the spectrum into clear absorption and continuum bands is not so obvious, but useful indices can still be defined. In contrast, the absorption in the 4000Å – 4100Å section is very great and several very strong Fe I lines are present. A number of lines of other species are also present. Even potential comparison bands suffer from appreciable amounts of absorption.

The broad regions of the spectrum in which the provisional indices lay were chosen to be the regions used for the metallicity estimation techniques. These three regions are 4025Å – 4090Å, 4500Å – 4690Å and 4870Å – 5000Å. An extensive grid of synthetic spectra was computed in these regions, as was described in detail in Section 2. These spectra were then available for the final choice of indices.

### 3.5 The selection of a set of indices for use in the analysis techniques

Once it had been established from the study of synthetic spectra what types of indices would be appropriate for the analysis of observed spectra (Section 3.3), it was necessary to determine whether a single set of indices would be appropriate for all stars of interest. The strength of absorption from the neutral iron group lines, the species of greatest concern, varies strongly with the atmospheric parameters of the stars under study. The sensitivities of indices to stellar parameters will consequently vary across the parameter range over which the analysis techniques are intended to be used (effective temperatures between 4800 and 6300K, metallicities between [Fe/H] = +0.3 and −2.0).

Figure 4 shows the 4900Å – 5000Å section of the spectrum of a solar metallicity K0 V star for a resolution of 1.0 Å (full-width at half-maximum intensity). This shows some of the strongest line absorption which is likely to be encountered among the stars of interest here. The sensitivities of regions within this wavelength range to metallicity can be judged from Figure 5a which shows the differences in intensity between the spectrum of Figure 4 and that of a similar but slightly metal-poor star (the difference between [Fe/H] = 0.0 and −0.5). Similarly, the sensitivities of regions to surface gravity are displayed in Figure 5b, which represents changes in intensity resulting from a change in $\log_{10}(g\,/\,\mathrm{cm\,s^{-2}})$ from 3.0 to 4.0. The same conclusions relating to index selection are obtained from this as were found from the study of the spectrum of a solar metallicity G0 V star in Section 3.3 .

In contrast, Figures 6 and 7a,b present analogous data for the spectrum of a metal-poor

G0 V star, as an example of one of the weakest absorption objects of interest. These may be compared with the K0 V results.

In general, the regions of the spectrum of a solar metallicity K0 V star which are most sensitive to metallicity are also the regions most sensitive in the metal-poor G0 V spectrum. Thus only one set of abundance indices was necessary for all stars and that they could be chosen from a single set of sensitivity plots. There are some differences in the gravity sensitivities of the spectra of Figures 5b and 7b – the metal-poor G0 V spectrum is comparatively insensitive to surface gravity – but such a star is an extreme example. Therefore a single set of gravity indices was adopted.

The spectrum of a solar metallicity G5 V star was chosen for the selection of the indices, since the degree of absorption is more typical of the stars of interest than is either of the two extreme examples considered above. By identifying the regions of high and low sensitivity to each of metallicity and gravity, the flux bands required to define the indices could be selected. The differences in intensity between spectra having [Fe/H] = 0.0 and −0.5, or $\log_{10}(g\,/\,\mathrm{cm\,s^{-2}}) = 3.0$ and 4.0 are presented in Figures 8 to 13.

It is possible to define three types of index, depending on the kind of spectral features the flux bands contain. They are:

 i)   iron abundance indices largely insensitive to surface gravity;
 ii)  iron abundance indices showing a sensitivity to gravity through the damping wings of strong lines;
 iii) ionic gravity indices (sensitive to gravity) measuring the relative absorption of ionic and neutral features.

The different relative sensitivities to metallicity and gravity of these indices allows the possibility of determining these two parameters.

Metallicity indices require absorption flux bands which measure iron line absorption and comparison bands which are insensitive to abundance. A list of suitable bands was compiled using the sensitivity data presented in Figures 8 to 13. The bands are ideally as wide as is possible whilst not including unnecessarily large amounts of nearby insensitive regions; at very least they had to have widths considerably in excess of the resolution of the spectra they were intended to analyse. If feasible, all bands were kept wider than 3.0 Å, with none narrower than 2.5 Å. Once a band had been selected from a study of the graphs, the Moore *et al.* (1966) solar line list was consulted to determine whether it was badly contaminated by species other than Fe I. If the absorption contributed by these other species was so large that it might adversely affect the ability to measure a true iron abundance, the band was rejected. Comparison indices were selected in a similar manner, except that they were based on regions which have small sensitivities to each of metallicity and gravity.

Ionic gravity indices were defined so that they compare ionic absorption features with lines from neutral species. Ideally these would measure the ratio of the fluxes in bands containing Fe II lines to others containing Fe I lines of similar strength, thus reducing the abundance dependence of such indices. However, this was not always possible. The sensitivity of these ionic lines to gravity depends on equivalent width and begins to saturate for very strong lines (e.g. Gray, 1976). Consequently, weak lines contribute important information. Unfortunately Fe II lines suitable for defining ionic indices are not common and it proved necessary to supplement them with ones of Cr II and Ti II, comparing them with Cr I and Ti I lines if available in order to reduce any influence from Cr and Ti abundances. This, again, cannot always be achieved and comparison bands had often to be chosen only because they are insensitive to gravity. Other species produce lines which are gravity sensitive, including a number of otherwise obscure ionic species of relatively low abundance. Flux bands should be chosen to avoid these features if possible.

The different wavelength regions synthesised were searched in turn for suitable flux bands. The 45 flux bands chosen as the bases of metallicity indices are listed in Table 6. Those selected for use in gravity indices, 42 in number, are presented in Table 7. There are 80 different bands in total. A set of 11 metallicity and metallicity-gravity indices have been defined from these and are listed in Table 8. A set of five ionic gravity indices is given in Table 9.

Note that the indices have been chosen for their sensitivity to iron abundance, rather than to some general metallicity parameter. In contrast, photometric metallicity estimates use indicators which measure the combined blanketing effects of spectral lines of many species. An advantage of the techniques developed here is that they derive true iron abundances.

# 4 IRON ABUNDANCE DETERMINATIONS FROM LOW-SIGNAL SPECTRA

## 4.1 The basic analysis methods

Iron abundances are determined by comparing the observed values of the abundance and surface gravity indices of Tables 6 to 9 with those of synthetic data corresponding to the known effective temperature of the star. The different sensitivities of the individual indices to abundance and gravity allow a solution for the two parameters to be obtained. Index values may be calculated from an observed spectrum. Similarly, theoretical indicator values can be computed from each of the synthetic spectra. Interpolation to the temperature of the star provides a set of synthetic index values at each point in a grid of abundances and gravities. These can be compared directly with the observed index data to provide an abundance–gravity relation from each index. These relations in turn allow a solution for the two parameters to be found.

As discussed above, 80 flux bands were chosen as the basis of the metallicity and gravity sensitive indices. Therefore, a set of 80 fluxes must be calculated from the synthetic spectra for each of the 100 stellar models, a total of 8000 values. The fluxes in the narrow wavelength bands are somewhat sensitive to the resolution of the spectrum as a result of the extent to which features at the edges of the bands are included. In order to allow the analysis methods to be used for the reduction of spectra of differing resolutions, flux data were computed appropriate to a range of resolution from 1.0 Å to 2.5 Å full-width at half-maximum in 0.1 Å intervals. Each theoretical spectrum was convolved with a Gaussian profile to represent instrumental effects, assuming a constant resolution across the spectrum. Residual fluxes were calculated by integrating the residual intensities over wavelength across each flux band. Each set of 8000 synthetic fluxes corresponding to a particular resolution was stored in a separate computer data file; the analysis methods can be tailored to the resolution of a particular observed spectrum by the choice of the synthetic flux data file.

The present techniques were developed to analyse spectra obtained from fibre-fed spectrographs. A difficulty with such systems, which complicates comparison of synthetic and observed fluxes, is the presence of scattered light in the spectrograph and detector, adding a locally-variable background signal to the reduced spectra. Care is essential to correct reliably for this scattered light. For the spectra of relevance here, the optimum sky background subtraction algorithm applies a local scattering correction (see Wyse and Gilmore 1992a), obviating the need for any special processes to handle scattered light in the abundance analysis. Therefore no correction was applied to the synthetic fluxes to account for any scattered light.

A set of index values can be computed from the synthetic fluxes. These 16 synthetic indices may be calculated using Equation 12 for each of the 100 different stellar models, giving a grid of 1600 data values. Therefore each of the 100 points in the temperature – gravity – metallicity ($T_{eff}$ – log $g$ – [Fe/H]) parameter space has a set of 16 index values associated with it.

Provided a temperature can be found from the star independently of these analysis methods, for example using photometric measurements, the problem of solving for the basic stellar physical parameters is greatly eased. This instantly provides one of the three stellar parameters which are used here to describe the physical nature of the star under study. Given the effective temperature, it is possible to calculate synthetic index values for each of the 16 indices of interest at each point in a grid in a metallicity – surface gravity plane corresponding to the temperature of the star. This may be accomplished by interpolation over temperature for each set of points in the $T_{eff}$ – log $g$ – [Fe/H] parameter space defined by a pair of gravity and metallicity values. The three-dimensional $T_{eff}$ – log $g$ – [Fe/H] parameter space therefore collapses to a two-dimensional [Fe/H] – log $g$ space, with 16 index values being associated with each of the 20 points in the plane. The cubic spline interpolation routines of Press *et al.* (1986) were used to compute the synthetic index values for the specified temperature.

The value of an observed index is determined, in the absence of errors in the measurement procedure, by the atmospheric parameters of a star. Conversely, if the value of an index is known, the possible range of combinations of the parameters is restricted. If the temperature is also known, the possible ranges of the remaining parameters are constrained even further. In the case adopted here where only three parameters are used to characterise the spectrum of the star, once both the temperature and an index value are available, a relation between the surface gravity and the metallicity may be found for that particular index. It becomes necessary therefore to determine the metallicity – gravity relations for each observed index numerically.

Once the effective temperature of a star is known, the value that a particular index may have is constrained to lie in a surface in the three-dimensional index – metallicity – gravity parameter space. This is illustrated in Figure 14. If the value of the index appropriate to that star is determined from observations, the line of intersection between the surface and the plane defined by the observed index can be found. This represents the [Fe/H] – log $g$ relation appropriate to the observed index and temperature of the star.

A set of computer routines was developed which implemented these principles, producing a different [Fe/H] – log $g$ relation from each index. Consequently, a set of 16 relations was derived. These data were used to solve for the iron abundance and, if possible, for the gravity.

## 4.2   A test of the analysis methods using noiseless synthetic spectra

Initially, tests involved noise-free synthetic data using the spectra of Section 2. The [Fe/H] – log $g$ relations of the indices intersected in a small region in the abundance – surface gravity plane. The scatter about the intersection point is due to numerical errors and its effect is very small– a few hundredths in the [Fe/H] parameter at most – and has a negligible effect on abundance results. However, these tests did not represent the analysis of genuine spectra realistically. All the spectra generated have, by definition, sets of stellar parameters which lie at the grid points in the temperature – abundance – gravity parameter space used to calibrate the reduction methods. Consequently, the extent to which the analysis software must perform interpolation when deriving the [Fe/H] – log $g$ relations is reduced.

To perform a more rigorous test, a set of synthetic spectra was computed having the atmospheric parameters $T_{eff} = 5200\,K$, $\log_{10}(g\,/\,\mathrm{cm\,s^{-2}}) = 4.2$, [Fe/H] $= -0.3$, and $\xi_{micro} = 1.5\,\mathrm{kms^{-1}}$. The data covered the same three wavelength ranges as the observational data which were later used to test the techniques further (see Jones, Gilmore and Wyse 1995), namely 4025Å – 4090Å, 4500Å – 4690Å and 4875Å – 4930Å. Of the original 16 abundance and gravity indices, 13 are found within these wavelengths. It was these 13 indices which were employed in the test. The three separate sections of spectrum were joined and were broadened to a 1.0 Å full-width at half-maximum resolution.

The artificial spectra were analysed using reduction procedures identical to those required for genuine observed spectra. The [Fe/H] – log $g$ relations were calculated successfully for all 13 indices. The results, in the form of the metallicity – gravity plot, are presented in Figure 15. A clear intersection point is found at the correct values of both [Fe/H] and log $g$. That this point is so well defined is a confirmation of the numerical accuracy of the analysis software.

The relations in Figure 15 are labelled according to the type of index from which they were derived. The relations belonging to those abundance indices which were selected for their sensitivities to metallicity alone, intended to be the primary source of [Fe/H] information, are indeed found to have the large gradients expected. In contrast, the metallicity – gravity indices have relations which are much shallower, allowing the intersection point to be identified without ambiguity. The remaining relations, measuring the ionic to neutral species equilibrium, permit a confirmation of the value of the surface gravity of the star.

The test spectrum considered above represents a comparatively late-type star which is only slightly metal-poor. Those spectral lines which contribute most to the absorption measured by the indices are therefore relatively strong. In order to assess the performance of the analysis techniques for stars which have weaker lines, it was necessary to generate a second test spectrum. This was given a set of stellar parameters appropriate to a hotter but more metal-poor star. The parameters chosen were $T_{eff} = 5850\,K$, $\log_{10}(g\,/\,{\rm cm\,s^{-2}}) = 3.20$, [Fe/H] $= -1.20$, and $\xi_{micro} = 1.5\,{\rm kms^{-1}}$. Although the intersection point of the [Fe/H] – log $g$ relations is not so well defined as in Figure 15 because of numerical errors during the analysis, it is still possible to determine the values of [Fe/H] and log $g$ accurately.

## 4.3 A test of the analysis technique using synthetic spectra with added noise

The analysis techniques are designed to deal with low-signal data. Therefore a more robust test is achieved using synthetic spectra with added noise, with characteristics similar to the noise in the intended observational data. Hence, normally-distributed noise was added to the intensities in each pixel of the test spectra of Section 4.2 using a nominal signal-to-noise ratio of 10 in 0.75 Å wide wavelength pixels. The metallicity – gravity relations obtained are shown in Figure 16 and 17. Errors are clearly a problem, with the intersection point lost in both cases. In Figure 16 it is possible to find an intersection region, but it is ill-defined and choosing the best ([Fe/H],log $g$) point becomes a relatively subjective process. An indication of the internal errors arising during the selection of a best point can be made by attempting to identify its position a number of times and calculating the scatter in the results. This method gave mean values (and standard deviations) of $-0.1$ ($\sigma = 0.1$) in [Fe/H] and of 3.9 ($\sigma = 0.2$) in log $g$ for the data of Figure 16. These compare favourably with the true parameter values ($-0.3$ and 4.2), but it must be emphasised that these errors will not be typical of an analysis of an observed spectrum. Temperature uncertainties will add further errors to the metallicity estimates made from observed data. The standard deviations about the mean results represent estimates of the accuracy of the eye estimation process alone; the noise induced errors may be significantly larger. Nevertheless, the basic methods which have been established to analyse stellar spectra appear to work adequately for this example.

The results obtained for the second test spectrum (Figure 17) are poorer. The scatter of the [Fe/H] – log $g$ relations is so large that no intersection region can be defined. As a consequence it is impossible to estimate a surface gravity, and only a very rough abundance estimate can be made, little more than a classification of the star as being metal-poor. However, of equal concern is the loss of some data : only 9 relations are found out of the 13 which should be present. This is a consequence of the rejection of data by the interpolation routines because excessive extrapolation would have been necessary. This is due to large random errors in fluxes, which alter the observed index values to such an extent that they lie outside the range of the synthetic index data. The loss constraints prevents a meaningful average metallicity from being calculated, even given an assumed gravity. Useful results are therefore not possible using these techniques, for stars which show as little line absorption as that of the second test spectrum, if the observational data are as poor as the artificial data considered here. The analysis methods in the form in which they have been described hitherto break down. The techniques must be modified if metallicities are to be found from stellar spectra of this kind.

### 4.4 Introducing compound indicators

The indices selected in Section 3 fall into one of three categories, depending on the nature of the spectral features they measure and, consequently, their sensitivities to stellar parameters. Although any two indices may be defined in terms of fluxes in two very different regions of the spectrum, they will have similar sensitivities to metallicity and gravity provided that they are from the same category. They therefore will produce similar relations in the [Fe/H] – log $g$ plane, as is confirmed by inspection of Figure 15. In principle it is possible to combine several indices from the same category to form a new indicator which would show sensitivities to stellar parameters which are typical of the individual indices. Such an indicator composed of a number of individual indices will, for convenience, be called a compound indicator.

The major advantage of using compound indicators is that the relative errors in the indicators used for the analysis are reduced. Reducing the errors has a beneficial effect during the use of the interpolation routines in the analysis program. With smaller relative errors it is less likely that data for a particular indicator will be rejected on account of the need to use unsafe extrapolation. As a consequence, more data will be retained for the metallicity estimation.

Another significant advantage resulting from the use of compound indicators concerns the estimation of parameters from the [Fe/H] – log $g$ relations. Whereas the use of single indices can produce a large number of relations which are distributed over a wide area in the metallicity–gravity plane, compound indices produce a smaller number which are less scattered. The parameter estimates are therefore likely to be more reliable.

We choose to define a compound indicator as the weighted sum of the single index values, normalising the result to a value of unity in the absence of absorption. The weighting factors are chosen according to the expected error in each index. An extra weighting factor can be introduced to give less importance to some indices, such as those which lie in regions of the spectrum where the continuum intensities in the raw spectra are lower and fewer counts are available in each pixel.

A compound indicator is defined in terms of single indices $I_1$, $I_2$, ..., $I_{N_I}$, as

$$C \equiv \sum_{j=1}^{N_I} \omega_j \, I_j \quad , \tag{15}$$

where $C$ is the compound indicator, $I_j$ is the value of the $j^{th}$ index used to define the compound indicator, $\omega_j$ is the weighting factor given to the $j^{th}$ index, and $N_I$ is the number of individual indices used. Neglecting the possibility of adding extra weighting, the weighting factors will be defined so that

$$\omega_j \propto \frac{1}{(\text{ error in } j^{th} \text{ index })^2} \quad . \tag{16}$$

The constant of proportionality allows for the normalisation of the compound indicator. The error in the individual indices can be estimated from Equation 14, on using the approximation that the residual fluxes in each band are close to unity. Substituting for the error gives

$$\omega_j = k \frac{(\sum_{i=1}^{N_{Aj}} \Delta\lambda_{A\,ij}) \, (\sum_{i=1}^{N_{Cj}} \Delta\lambda_{C\,ij})}{\sum_{i=1}^{N_{Aj}} \Delta\lambda_{A\,ij} + \sum_{i=1}^{N_{Cj}} \Delta\lambda_{C\,ij}} \quad , \tag{17}$$

where $\Delta\lambda_{A\,ij}$ is the wavelength width of the $i^{th}$ absorption band of the $j^{th}$ individual index, $\Delta\lambda_{C\,ij}$ is the wavelength width of the $i^{th}$ comparison band of the $j^{th}$ individual index, $N_{Aj}$ is the number of absorption bands used to define the $j^{th}$ index, and $N_{Cj}$ is the number of comparison bands used to define the $j^{th}$ index. On substituting for $\omega_j$ from Equation 17 into the expression for the compound indicator in Equation 15,

$$C = k \sum_{j=1}^{N_I} \frac{l_{Aj} \, l_{Cj}}{l_{Aj} + l_{Cj}} \, I_j \quad . \tag{18}$$

where the terms $l_{Aj}$ and $l_{Cj}$ are abbreviations for the summations of the band widths so that

$$l_{Aj} \equiv \sum_{i=1}^{N_{Aj}} \Delta \lambda_{Aij} \quad , \text{ and } l_{Cj} \equiv \sum_{i=1}^{N_{Cj}} \Delta \lambda_{Cij} \quad . \tag{19}$$

The compound indicator may be normalised to a value of unity in the absence of absorption by selecting a value of $k$ so that $C = 1$ when $I_j = 1$ for all $j$. On substituting for $k$, Equation 18 becomes

$$C = \frac{\sum_{j=1}^{N_I} \frac{l_{Aj} \, l_{Cj}}{l_{Aj} + l_{Cj}} \, I_j}{\sum_{j=1}^{N_I} \frac{l_{Aj} \, l_{Cj}}{l_{Aj} + l_{Cj}}} \quad . \tag{20}$$

This is the expression which will be used to define compound indicators when no additional weighting is used.

If instead additional weighting factors are used, the compound indicator should be defined as

$$C \equiv \sum_{j=1}^{N_I} \omega_j \, w_j \, I_j \quad , \tag{21}$$

where $w_j$ is the additional weighting factor given to the $j^{th}$ index. The $\omega_j$ weighting factors however remain unchanged. Defining the factors $\omega_j$ as in Equation 16, and requiring that $C = 1$ when $I_j = 1$ for all $j$, we obtain

$$C = \frac{\sum_{j=1}^{N_I} \frac{w_j \, l_{Aj} \, l_{Cj}}{l_{Aj} + l_{Cj}} \, I_j}{\sum_{j=1}^{N_I} \frac{w_j \, l_{Aj} \, l_{Cj}}{l_{Aj} + l_{Cj}}} \quad . \tag{22}$$

It is now possible to define and calculate compound indicators using Equation 20, or Equation 22 if additional weighting factors are used. This option has been incorporated into the analysis software. The [Fe/H] – log $g$ relations are returned for each compound indicator and can be used to select a best intersection point as in the case of the single indices discussed above.

Six compound indicators have been selected. They are listed in Table 10. Only one of the six uses the additional weighting factors. This is to give lesser importance to an index defined in the 4025Å – 4090Å region of the spectrum where the signal is relatively weak in the test spectra of paper II and for the data of the faint F/G star survey. Three compound indicators are sensitive to metallicity but not to gravity, one is sensitive to both (because of damping effects of very strong iron lines) and two attempt to measure gravity form the ionic to neutral species equilibrium.

## 4.5 Testing the analysis technique using compound indicators

The test spectra which were used in Section 4.2 to investigate the performance of the analysis techniques and software with conventional indices were also suitable for use in assessing their operation with compound indicators. A reduced set of compound indicators was defined, as discussed above, covering the restricted wavelength ranges of the synthetic test spectra. The behaviour with noise-free spectra is illustrated in Figure 18. The intersection points of the [Fe/H] – log $g$ relations are clearly defined.

The results of analysing the two noise-added test spectra using compound indicators are presented in Figures 19 and 20. It is clear that the scatter in the [Fe/H] – log $g$ relations for the first test spectrum is greatly reduced when the compound indices are used in preference to the single indices. Estimation of the best ([Fe/H], log $g$) point is therefore simplified. The values of the metallicity and the gravity obtained are not significantly altered by the change in analysis technique. However, one compound index ionic gravity relation is still lost because extrapolation is not performed. The second test spectrum, having considerably weaker absorption, gives a large scatter in this plane (Figure 20). The first compound index relation is lost, losing some important metallicity data, but the other five relations remain. This represents an improvement over the single index results in that more data are kept. It still remains impossible to estimate a surface gravity and even a mean metallicity is difficult to find because of the scatter.

The use of compound indicators represents a substantial improvement over single indices, particularly in terms of the retention of data when errors are large. However, there are occasions, such as in the analysis of stars having weak absorption, when no attempt can be made to find metallicity and gravity data using eye estimates, at least for spectra as noisy as the two synthetic examples considered here. There is little alternative but to adapt the analysis methods further and to resort to some appropriate assumptions in order to be able to estimate metallicities for these cases.

### 4.6    Assuming a value of the surface gravity

It is usually desirable to be able to estimate surface gravities in addition to iron abundances for stars. This can be particularly true for studies of field stars in which it is likely that the sample will include both dwarfs and giants, in order to be able to account for luminosity or evolutionary effects. However, the primary objective of the analysis methods discussed here is to derive metallicity information, with gravities being only of secondary interest. Thus we assumed an input value of the surface gravity, when required. There may often be evidence that stars of a given luminosity class will dominate a sample of objects, in which case it becomes reasonable to assume an appropriate gravity for that type of star. This is indeed true of the faint F/G star survey for which these techniques were developed, and which is dominated by dwarfs.

It is possible, in principle, to estimate an iron abundance on the assumption of a surface gravity by using either the sixteen indices of Tables 8 and 9 or the six compound indicators of Table 10. However, such an approach would suffer the problem that the error in an index or indicator value may be so large that interpolation, or more correctly extrapolation, by the analysis software within the synthetic data could not reliably reach the observed values. Consequently, no [Fe/H] – log $g$ relation could be returned for some indices or indicators, reducing the available metallicity information. The final method selected for the determination of abundance using an assumed surface gravity involved the adoption of a single compound indicator based on all abundance-sensitive indices. The selection and use of such comprehensive compound indicators will now be discussed.

### 4.7    Abundance determinations using comprehensive compound indicators

To perform abundance analyses of particularly noisy spectra, a compound indicator, of the type discussed above, was defined using all the metallicity-sensitive indices of Table 8. When combined with an assumed surface gravity, the data produced by this indicator can often provide a [Fe/H] result free of the problems caused by the selective loss of indicator data which were described in the previous section. It is this indicator which was adopted as the main method of deriving abundance information from spectra for which the six indicator cursor method failed.

The single, comprehensive compound indicator was defined from all the available metallicity indices. Some of these indices are sensitive to iron abundance but not to surface gravity, while others, measuring absorption by strong Fe I lines, show a sensitivity to both abundance and gravity. In order to reduce the gravity sensitivity of the compound indicator, and consequently to reduce the error in a [Fe/H] result produced by an inappropriate choice of a surface gravity, the abundance-only sensitive indices were given weighting factors in Equation 22 of $w_j = 2$, while the abundance and gravity sensitive indices were given weights of $w_j = 1$.

To allow some statistical assessment of the importance of the assumed surface gravity in abundance results for samples of stars, a second comprehensive compound indicator was defined. This used the indices of Table 8 which were sensitive to metallicity only, rejecting those which showed some appreciable gravity sensitivity. Being made of only seven indices, as opposed to the eleven of the all-metallicity indicator, this metallicity-only indicator will suffer more severely from noise-induced errors. Therefore the all-metallicity results are to be preferred to those of the metallicity-only indicator for the study of particularly noisy spectra.

Extensive tests and evaluations of the performance of the comprehensive compound indicators have been carried out using synthetic spectra. They have established that the indicators provide an efficient and reliable method of deriving abundance information from particularly noisy spectra. These studies show that the random error in the derived iron abundance is highly sensitive to the stellar metallicity and temperature, as well as to the signal-to-noise ratio of the spectrum. The random error in [Fe/H] is less than $\pm 0.2$ for solar metallicity G-type stars of all temperatures for signal-to-noise ratios as low as 10 in each 1.0Å wide resolution element. The error increases with decreasing line absorption strength (increasing temperature or decreasing metallicity). It is still smaller than $\pm 0.3$ for signal-to-noise ratios of 20, for G dwarfs having [Fe/H] $= -1.5$. This confirms that the indicator defined from all metallicity-sensitive indices is capable of supplying useful iron abundance data from even very low-signal spectra.

Systematic errors could cause a distortion of the scale of [Fe/H] results, which could adversely affect any conclusions drawn about stars and stellar populations, based on abundances obtained using the present analysis techniques. Errors caused by deficiencies in the computation of synthetic spectra were discussed in Section 2. Whether a recalibration of the abundance scale is appropriate is discussed in Paper II. The possibility that errors may be introduced by inadequacies in the wavelength scale calibration methods and by the adoption of a slightly inappropriate resolution for the synthetic spectra are considered in Section 5 below. Systematic errors introduced by the performance of the comprehensive compound indicators are discussed here.

The motivation behind the adoption of the all-metallicity indicator was to overcome the frequent failure of indices or indicators with a more restricted total wavelength coverage to derive [Fe/H] – $\log g$ relations when the noise-induced error in the index values were large. Although the use of the single comprehensive indicator represents a considerable improvement in that the likelihood of such failures is substantially decreased, there are occasions when even the comprehensive indicator fails to produce [Fe/H] – $\log g$ relation data. The potential danger that this introduces is that the failures will be selective, in that they will occur preferentially and will cause a systematic error in the abundance results, overestimating the abundances of metal-poor stars. Similarly, there will be a tendency to underestimate the abundances of metal-rich stars, although the effect will be considerably less important than in the metal-poor case, because the importance of random errors in the analysis of spectra showing strong Fe I line absorption is much less than in those experiencing weak absorption.

A second noise-induced systematic error in the [Fe/H] results is to be expected, but operating so as to underestimate the abundance, in contrast to the effects discussed above. It is caused by the loss of sensitivity of the abundance indices at low metallicities. Because there is only a small amount of absorption in the spectrum at such low metallicities, a small change in the abundance of a star would produce only a small change in the detected absorption and hence only a small change in the index and indicator values. These 'degeneracy effects' are also found in traditional photometric metallicity indicators such as the $\delta(U - B)_{0.6}$ parameter of the UBV broad-band system or the $\delta m_1$ parameter of the intermediate-band uvby$\beta$ system. If the errors in the observed indicator measurements are large, the errors in the abundance estimates will be correspondingly large. However, if the sensitivity of the index changes significantly at these values, the abundance error distribution will become asymmetric even if the distribution of the indicator errors is symmetric. The asymmetric distribution will produce a systematic shift in the mean [Fe/H] result. Further investigations were required to establish the size of this effect in the abundance analyses described here; these are described in Section 5.1 below.

# 5 THE EXPECTED ERRORS IN THE IRON ABUNDANCE RESULTS

The iron abundance determination methods are designed to be used for the analysis of low-signal spectra. As the effects of noise in the observed data are expected to be the limitation on the accuracy of the iron abundance results, we wish to understand the scale of these errors as a function of the signal-to-noise ratios of the spectra and of the stellar physical parameters. Other sources of error such as the use of inappropriate effective temperatures or surface gravities for the stars under study are investigated below.

## 5.1 The effects of noise on an iron abundance analysis

The error in an iron abundance measurement for a given signal-to-noise ratio of a spectrum will be a strong function of the degree of metal-line absorption, and hence of stellar effective temperature and metallicity. It is important to establish a detailed understanding of how the size of the error in [Fe/H] caused by noise varies as a function of various important stellar physical characteristics and observational parameters. Numerical modelling has been used to investigate the functional dependance of this error on the signal-to-noise ratio, temperature and metallicity. The comprehensive compound indicator sensitive to both metallicity and gravity (the 'all-metallicity' indicator) is intended to be the primary source of abundance results for low-signal spectra. Consequently, only the analyses performed using this indicator will be considered.

A numerical method was adopted to determine the detailed variation of the errors in [Fe/H] with appropriate parameters. A Monte Carlo simulation of the effects of random errors in the intensity data of a spectrum was carried out. Repeated analyses were performed on noise-added synthetic spectra, building a set of [Fe/H] measurements for each particular type of spectrum. This method had the advantage that the detailed characteristics of the [Fe/H] error distribution could be found if required, including the effects of both random and systematic errors.

A resolution of 1.0Å full-width at half-maximum was adopted, being the resolution for which the techniques are optimised. The noise-added spectra were analysed using the all-metallicity comprehensive compound indicator, as amended to use the restricted 4025Å – 4930Å wavelength region of the observational data of Paper II. Reasonable estimates of the size of the error could be obtained with as few as 10 different analyses on a given spectrum (using different random noise for each) for ratios of $R_{S/N\,0.5\text{Å}} = 30$ and $50$. However, up to 30 attempts were required for spectra having the other ratio values in order to be able to estimate the error with any reliability. The standard deviation of the individual [Fe/H] results about their mean gave an estimate of the [Fe/H] error for each spectrum, and allowed a table of error data to be built up. Because the uncertainties in the individual error estimates were likely to be relatively large, the data were smoothed by fitting a functional representation through the points. The smoothed results are presented in Table 11, expressed for signal-to-noise ratios equivalent to 1.0Å wide pixels. It is clear from Table 11 that the analysis techniques work well for lower noise spectra. However, the noise-induced [Fe/H] increases rapidly with ratios below $R_{S/N\,1\text{Å}} = 20$, with little useful information being provided below $R_{S/N\,1\text{Å}} = 10$.

The numerical simulations were repeated for spectra having the coarser resolution of 2.1Å. This is the quality of the observational data used to test the analysis techniques (Jones, Wyse and Gilmore 1995). The noise-induced errors were typically 10% – 30% larger than the 1.0Å case for similar spectra. This larger error is as expected given the loss of sensitivity of indices to metallicity, when the resolution is degraded.

The possibility that noise in the intensity data of a spectrum might introduce systematic errors into [Fe/H] results, in addition to random errors, was briefly discussed above. These systematic errors could be caused by the variation in the sensitivity of index or indicator values to abundance being significant over the scale of the uncertainties in the values. The resultant [Fe/H] error distribution would become distorted and its asymmetry could produce a general shift in the [Fe/H] results to lower values. Equally the selective failure of analysis attempts close to the boundaries of the grid of synthetic spectra in the stellar physical parameter space could introduce other systematic effects. These systematic errors could lead to a distortion of the abundance scale.

A measurement of the likely systematic error in a [Fe/H] result was made by determining the drift of the mean of the distribution of [Fe/H] values as a function of signal-to-noise ratio. The number of analysis attempts which had to be made to obtain a reliable estimate of the systematic error was, however, larger than that needed to make the random error estimates discussed above. The simulations were carried out for the 2.1Å resolution of the observational data analysed in Gilmore, Wyse and Jones (1995). As expected, the systematic error was found to be a strong function of the noise in the spectrum. For signal-to-noise ratios $R_{S/N\,1\text{Å}} > 15$, the systematic error is generally smaller than 0.1 in [Fe/H]. Even at $R_{S/N\ 1\mathring{A}} = 10$, the systematic error is typically less than 0.15 for [Fe/H] > −1.0. For these very noisy spectra the systematic error is generally significantly smaller than the random error. On this basis, it is reasonable not to recalibrate the [Fe/H] scale to correct for noise-induced systematic abundance errors.

### 5.2 The errors in the iron abundance caused by errors in stellar temperatures

The abundance analysis techniques require a measurement of the temperature of a star to be available before an attempt can be made to solve for its abundance and gravity. Such temperatures can be determined by the use of an appropriate broad-band or intermediate-band photometric colour index (see Jones, Wyse and Gilmore 1995 for a discussion of the calibration of the (V−I)$_C$ index). Any error in the temperature specified for the star will produce a corresponding error in the iron abundance result. These effects were investigated for the comprehensive compound indicator which is sensitive to both metallicity and gravity defined in the 4025Å – 4930Å region discussed in Section 4.4. The simulations were performed for resolutions of 1.0Å full-width at half-maximum intensity.

Table 12 lists the random errors in the effective temperature results which would be produced by various errors in the (V−I)$_C$ colour index of a star. The size of the error is dependent on the temperature of the star and is therefore given for several different values of $T_{eff}$. A systematic error may also to be present in the temperature results, caused by general uncertainties in stellar temperature scales introducing an error into the original (V−I)$_C$–$T_{eff}$ calibration. The errors in [Fe/H] produced by the photometric uncertainties are presented in Table 13.

### 5.3 The errors in [Fe/H] introduced by assuming an inappropriate surface gravity

The abundance analysis techniques are incapable of solving for both a stellar iron abundance and for a surface gravity if the spectrum is particularly noisy. In these instances, it is necessary to assume a surface gravity if an abundance result is to be obtained, as discussed in Section 4.

The errors in [Fe/H] which would be produced by an error of ± 1.0 in the assumed value of $\log_{10} g$ are presented in Table 14 for different types of star. These have been estimated from the [Fe/H]–log $g$ relations which are returned on performing abundance analyses of ap-propriate synthetic spectra. The data apply to the all-metallicity comprehensive compound indicator in the restricted 4025Å – 4930Å region.

The data of Table 14 suggest that uncertainties in assumed surface gravities can introduce significant errors into the [Fe/H] results obtained using the metallicity and gravity sensitive indicator when observing samples of stars having a mix of luminosity classes. The use of the gravity-insensitive indicator is therefore to be recommended for the analysis of spectra which are not so badly affected by noise. However, for very low signal data, there is little alternative to the use of the indicator which uses the greatest number of abundance-sensitive flux bands.

## 5.4 The errors in [Fe/H] caused by the use of incorrect microturbulence parameters

The abundance analysis techniques have not been designed to allow microturbulence parameters to be determined. A single microturbulence was adopted for all stars. This assumed value is intrinsic to the analysis techniques, being the value used to compute the synthetic spectra and the synthetic fluxes which are compared with the observed indicators. Should the observed stars experience different microturbulence, there will be a corresponding error in the [Fe/H] results. A systematic difference in microturbulence from the assumed value would lead to a systematic error in the abundance scale.

It is necessary to attempt to quantify the sizes of the [Fe/H] errors which would be caused by the use of an incorrect microturbulence. For this purpose, some synthetic spectra were computed for microturbulence parameters of $\xi_{micro} = 1.0\,\mathrm{kms^{-1}}$, in addition to the extensive grid having $\xi_{micro} = 1.5\,\mathrm{kms^{-1}}$, as was described in Section 2.2 . Analyses of these $\xi_{micro} = 1.0\,\mathrm{kms^{-1}}$ spectra provided information about the errors introduced by assuming an incorrect microturbulence parameter. These results are listed in Table 15. The errors in the [Fe/H] results which are produced by errors in the assumed microturbulences can be significant for some spectra. In contrast to the noise-induced [Fe/H] errors, they are at their greatest for cooler and more metal-rich stars.

## 5.5 Abundance errors caused by deficiencies in the wavelength scale calibration

Errors in the calibration of the wavelength scale of an observed spectrum will cause incorrect wavelength limits to be used for the flux bands which form the bases of the abundance indices and indicators. Some spectral features will be partly or even totally lost from flux bands, while absorption from other features will be introduced into them, causing an error in the [Fe/H] result.

Because the absorption flux bands of the abundance indices coincide with regions of the spectrum which experience strong line absorption, a wavelength error will most often shift the absorption flux band to a region of weaker absorption. Therefore, the value of an observed abundance index would in general be increased (towards the value it would have for weaker line absorption if the wavelength error were absent). This will produce a systematic error in the estimated iron abundance, with the abundance of the star of interest being underestimated.

Table 16 illustrates the significance of wavelength scale errors by presenting the error in [Fe/H] caused by a constant shift in wavelength for 1.0 Å resolution spectra. These data were generated by analysing synthetic spectra which had been displaced in wavelength by the constant shift in the table.

## 6 DISCUSSION

One may compare the present analysis techniques with methods established by other researchers for the reduction of low signal-to-noise spectra. The methods of Carney, Laird, Latham and Kurucz (1987), applied to a large sample of stars by Laird, Carney and Latham (1988), were designed to be used for the study of short sections (several tens of Ångström units wide) of high-resolution (about 0.2 Å) spectra. Direct pixel-by-pixel comparisons of observed spectra with a grid of synthetic data allowed a best metallicity to be found by minimising the sum of the squares of the residuals. Despite having narrow sections of spectrum, the use of the very strong Mg I, MgH and Fe I absorption features, together with the high resolution, allowed results of a very impressive accuracy to be obtained. They quoted typical signal-to-noise ratios for their survey spectra equivalent to 25–30 for 1.0 Å wide wavelength bins (signal-to-noise ratios will be expressed as the values appropriate to hypothetical 1.0 Å wavelength intervals, $R_{S/N\,1\text{Å}}$, to allow a direct comparison to be made between different techniques). For these data they estimated a noise-induced [Fe/H] error of ±0.12 . For spectra with $R_{S/N\,1\text{Å}} = 18$ , the error rose to ±0.17 . They stated that they found little dependence of the [Fe/H] error on metallicity, in sharp contrast to the techniques described here, due to their use of absorption features which are generally considerably stronger than the Fe I lines of this study. However, it is clear that for any light level, the two sets of techniques are competitive for moderately metal-poor stars, with the present methods being superior for solar metallicity stars but those of Carney *et al.* very substantially better for halo objects. Due to the different temperature sensitivities of the absorption features used by the two techniques, the Carney *et al.* [Fe/H] results will be more affected by a given error in the adopted effective temperature of a star (although Laird, Carney and Latham derived temperatures from several different photoelectrically-determined photometric colour indices, obtaining relatively accurate temperatures).

The significant difference between the Carney *et al.* metallicity determination methods and the present iron abundance measurement techniques is that they operate on very different types of observational data. The Laird, Carney and Latham (1988) spectra were obtained using échelle spectrographs, recording data for a single object at a time. In contrast, the present techniques are optimised for the analysis of intermediate-resolution spectra obtained using optical fibre multi-object spectroscopy, which allows many spectra to be recorded in a single exposure. The present technique can therefore be used in surveys of large numbers of very faint stars and represents an appreciable improvement in the efficiency with which metallicity information can be derived.

In contrast, the techniques of Cayrel, Perrin, Barbuy and Buser (1991a,b) were designed to be used for the reduction of intermediate-resolution spectra. Their methods are directly comparable to those discussed here, even though the data on which they were used had resolutions which, at 4.7 Å, were coarser than those of interest to this work. By minimising the sum of the squares of residuals between the spectrum of a star and sets of observed and synthetic templates, Cayrel *et al.* solved for effective temperatures and gravities as well as metallicities. Their method relies on various regions within the wavelength range under study (4780 Å –5300 Å) having different sensitivities to the three parameters to solve for them. They compared the observed spectra to the templates at each pixel, calculating the residuals at each. The need to reproduce the wavelength binning structure of each observed spectrum demanded that they performed a convolution and rebinning procedure on each synthetic spectrum for each abundance analysis attempt. Carney *et al.* were able to avoid this by using over-sampled data. An advantage of the Cayrel *et al.* techniques is that they solve for the stellar temperature, removing the need for an additional photometric measurement and an estimate of the reddening.

Cayrel *et al.* performed an error analysis for their abundance results. By analysing synthetic spectra to which random noise had been added, they estimated the errors in the measurements of each of the three stellar parameters. Their data for a 1.0 Å resolution can be compared directly with the results here. They considered spectra of dwarf stars having an effective temperature of $5000\,K$ and a signal-to-noise ratio of 40. The errors in the determined [Fe/H] parameters for stars having [Fe/H] $= 0.0$, $-0.75$, and $-1.75$ were found to be as small as 0.02, 0.02 and 0.04 respectively. Assuming that their quoted signal-to-noise ratios apply for wavelength bins having the width of the resolution element, the equivalent noise-induced errors of the present analysis techniques are 0.03, 0.05 and 0.09. The Cayrel *et al.* results appear to have a smaller noise-induced [Fe/H] error, particularly for metal-poor stars. However, an interpretation of their quoted [Fe/H] errors will be complicated by the fact they attempt to solve for temperature as well as metallicity, while the errors in the two parameters are strongly correlated.

Barbuy, Perrin and Cayrel (1990) attempted to use spectrophotometric metallicity and gravity indices to determine information from low-signal spectra of late-type stars in the 4780Å – 5300Å region. They obtained satisfactory results for a set of standard stars, but chose to to abandon the techniques in favour of those of Cayrel *et al.* Although conventional flux bands were used to measure regions of strong line absorption, the continuum level was defined by intensities at three 'pseudo-continuum' points. This continuum will therefore be very sensitive to the intensity errors at these points. Noise in the spectra might affect the metallicity results adversely.

Ratnatunga and Freeman (1989) used a cross-correlation method to determine metallicities of distant K giant stars from 2 Å resolution digital spectra. They defined two abundance-sensitive parameters which measured the size of the peak in the autocorrelation function of the spectrum of the star of interest. These parameters were calibrated against [Fe/H] and the dereddened (B−V) colour index using observational data for globular cluster stars (putting the [Fe/H] results on the globular cluster abundance scale). The parameters were defined in the 4000Å – 4410Å (containing strong metal and molecular absorption features, but with the G band data artificially removed) and 4920Å – 5440Å (including the Mg b feature, Mg H and Fe I absorption) wavelength regions respectively. The spectra had typical signal-to-noise ratios equivalent to $R_{S/N\,1\mathrm{\AA}} = 10$ and gave [Fe/H] errors of $\pm$ 0.15–0.30 comparable to, or slightly better than, the techniques presented here. Like those of Cayrel *et al.,* the methods of Ratnatunga and Freeman are suitable for use in multi-object spectroscopy.

It is clear that analysis methods exist for the determination of stellar metallicities from very low signal, intermediate-resolution spectra of late type stars. Despite distinct approaches, the techniques of Cayrel *et al.,* Ratnatunga and Freeman and of this work are able to provide metallicity data of broadly comparable accuracy, although the attempt of Cayrel *et al.* to solve for three stellar parameters may complicate their method's ability to operate at the lowest signal-to-noise ratios. Cayrel *et al.,* and Ratnatunga and Freeman chose to use the very strong absorption features around 5150Å– 5250Å (as did Carney *et al.*). The advantage of the methods of Carney *et al.,* Cayrel *et al.,* and Ratnatunga and Freeman over those presented here is that they use all the information in broad regions of a spectrum, whereas for these methods only selected narrow regions are employed. That the difference in accuracy is not great suggests that a selective approach does not necessarily sacrifice large quantities of important information; the wavelength regions which are overlooked, being only mildly sensitive to abundance, will not contribute greatly to an abundance result. Conversely, an appreciable advantage of such a selective approach is that it is possible to use only the regions of the spectrum for which it is believed that it is possible to compute synthetic spectra of reasonable accuracy, for example avoiding features having poorly determined atomic data in order to reduce systematic errors.

Perhaps the most obvious improvement which could be made to the techniques described in this paper would be an attempt to reduce the noise-induced errors in the [Fe/H] results. There are two distinct ways in which this could be achieved, both of which involve defining additional abundance index flux bands. Firstly, it would be possible to extend the synthetic spectrum grid to include wavelength regions not previously considered. In particular, within the 4000Å – 5000Å region, the 4420Å – 4500Å region (between strong CH absorption and the limit of the current synthetic spectra) and the 4690Å – 4845Å region (between the current cut-off and the H$\beta$ line) could allow several new indices to be defined. Alternatively, other wavelengths could be used if observational data could be obtained outside the 4000Å – 5000Å range.

A second way in which the numbers of abundance index flux bands could be increased would be by relaxing the criteria used to select indices. If absorption lines of species other than Fe I were also used (e.g. Ca I, Mg I, Ti I, Ni I) to define general metallicity indicators (as opposed to the iron abundance indicators of this work), it is likely that many new flux bands could be identified within the wavelengths of the existing synthetic spectra. Such general metallicity estimates would be most suitable for the analysis of very noisy spectra in order to reduce the noise-induced errors; for spectra of higher quality, where the noise-induced error is not of such importance as systematic errors, the existing iron abundance measurement methods could be used. It should be noted that most other techniques for the study of noisy spectra (e.g Carney *et al,.* 1987, Ratnatunga and Freeman, 1989, Cayrel *et al.,* 1991) derive general metallicities, which in the case of synthetic calibrations must be subjected to an additional, empirical recalibration to obtain [Fe/H] parameters.

Techniques similar to those presented here, using numbers of dedicated abundance indices, could be used to derive element ratio information from spectra of higher quality than those analysed here, enabling these data to be obtained rapidly and efficiently for large numbers of local dwarfs and possibly even for distant giant stars. These indices could measure large numbers of moderately strong lines or a smaller number of very strong features (*e.g.* the Mg b lines, the infrared calcium triplet). If molecular synthesis could be incorporated into the index calibrations, it should be possible to derive abundance ratio information from molecular bands even at low light levels, although a strong gravity dependence could complicate the analysis. Element ratios are important diagnostics in studying the properties of Galactic components (e.g. Wheeler, Sneden and Truran 1989; Gilmore and Wyse 1991; Wyse and Gilmore 1992b). If these methods could be pushed to low light levels, it would be possible to study element ratios in distant, unevolved stars, greatly extending their present use.

## 7 CONCLUSIONS

The advent of wide-field multi-fibre spectrographs has made possible large radial-velocity surveys of distant stars. While the general motivation for the design of these systems has primarily been the acquisition of kinematic data, we have established techniques which allow genuine iron abundances to be determined from such low-signal, intermediate-resolution spectra. Through the use of an extensive set of spectroscopic indices, they are capable of solving for the [Fe/H] parameter and surface gravity of F/G stars, requiring only an independent temperature estimate. For very low-signal data, the techniques have been adapted to provide an iron abundance corresponding to a specified surface gravity. The techniques have been designed to be operated with 1.0Å resolution data (full-width at half-maximum intensity), but can be applied to lower-resolution spectra. Calibrations have been performed for resolutions from 1.0Å to 2.5Å. The errors in the abundance results have been investigated in detail and the methods provide useful data from spectra having signal-to-noise ratios (in 1.0Å intervals) as low as 10. Higher signal data will produce correspondingly more accurate abundance data.

It is clear that, even in their present form, the analysis techniques are a powerful tool for determinations of metallicities of large numbers of faint stars. Unlike conventional high-resolution analyses, they can provide abundance information rapidly and efficiently. When combined with multi-fibre spectroscopy, they are able to study stars in sufficient numbers that statistical information about the metallicity and kinematic properties of stellar populations can be determined, at or beyond the magnitude limits even of photometric surveys. The practical application of these techniques and calibration using real data, together with a new effective temperature – colour relation, are described in Jones, Wyse and Gilmore (1995). The technique is precise – a small zero-point error of $\sim 0.1$ dex is calibrated by analyses of twilight sky spectra – and reliable, with the uncertainty being $\lesssim 0.2$ dex over the range of metallicities of the Galactic thin and thick disks. Gilmore, Wyse and Jones (1995) present the derivation of the thick disk iron abundance distribution from an *in situ* sample of F/G starswith magnitudes $V = 16$ to $18^m$ in high Galactic latitude fields, using metallicity data derived using these techniques.

## 8  ACKNOWLEDGEMENTS

The Center for Particle Astrophysics is supported by the NSF. RFGW acknowledges support from the AAS Small Research Grants Program in the form of a grant from NASA administered by the AAS, from the NSF (AST-8807799 and AST-9016226) and from the Seaver Foundation. Our collaboration was aided by grants from NATO Scientific Affairs Division and from the NSF (INT-9113306). GG thanks Mount Wilson and Las Campanas Observatories for access to their excellent facilities, as a Visiting Associate, during the early stages of this work. JBJ acknowledges financial support through a Science and Engineering Research Council studentship, and from the Institute of Astronomy, Cambridge. We wish to thank Dr. M. G. Edmunds for helpful discussions during this work.

## 10  FIGURE CAPTIONS

Figure 1. The Fe I damping enhancement factors of Simmons and Blackwell (1982, diagonal crosses), of Gratton and Sneden (1988, open circles) and of Gurtovenko, Fedorchenko and Kondrashova (1982, vertical crosses) plotted against lower level excitation potential. The solid line indicates the final enhancement–excitation potential relation adopted for the computation of the grid of synthetic spectra.

Figure 2. A comparison of synthetic and observed spectra of Arcturus. The 4938–4946 Å region of the spectrum of the moderately metal-poor K2 giant Arcturus is shown. The solid line indicates the synthetic spectrum computed with the oscillator strength data derived from the solar spectrum. The dashed line represents the observed spectrum of the star taken from the Griffin (1968) atlas.

Figure 3. The recalibration of the Moore, Minnaert and Houtgast equivalent width scale. The graph shows the equivalent widths presented by Moore, Minnaert and Houtgast (1966) for a number of absorption lines in the solar disc centre spectrum plotted against equivalent widths selected from published modern spectroscopic abundance analyses. The adopted recalibration relation is shown alongside the one-to-one relation. Diagonal crosses show data for Ti I lines from Blackwell, Booth, Menon & Petford (1987), open circles Cr I lines from Blackwell, Booth, Menon & Petford (1987), vertical crosses Fe I lines from Blackwell, Booth & Petford (1984), and asterisks strong Fe I lines from Blackwell & Shallis (1979)

Figure 4. The 4900–5000 Å region of the synthetic spectrum of a solar-metallicity K0V star. The graph shows the synthetic spectrum of a $T_{eff} = 5000\,\mathrm{K}$, $\log_{10}(g/\mathrm{cms}^{-2}) = 4.0$, [Fe/H] = 0.0 star. Its resolution is 1.0 Å (full-width at half-maximum). Line data are from the Moore, Minnaert and Houtgast (1966) list.

Figure 5a. The metallicity sensitivity of the 4900–5000 Å region of the spectrum of a solar-metallicity K0V star. The graph shows the difference in residual intensity produced by a decrease in metallicity from [Fe/H] = 0.0 to −0.5 in a $T_{eff} = 5000\,\mathrm{K}$, $\log_{10}(g/\mathrm{cms}^{-2}) = 4.0$ star as a function of wavelength. The resolution of the spectrum is 1.0 Å (full-width at half-maximum). Line data are from the Moore, Minnaert and Houtgast (1966) list.

Figure 5b. The gravity sensitivity of the 4900–5000 Å region of the spectrum of a solar-metallicity K0 IV star. The graph shows the difference in residual intensity produced by a change in surface gravity from $\log_{10}(g/\mathrm{cms}^{-2}) = 3.0$ to 4.0 in a $T_{eff} = 5000\,\mathrm{K}$, [Fe/H] = 0.0 star as a function of wavelength. The resolution of the spectrum is 1.0 Å (full-width at half-maximum). Line data are from the Moore, Minnaert and Houtgast (1966) list.

Figure 6. The 4900–5000 Å region of the synthetic spectrum of a metal-poor G0V star. The graph shows the synthetic spectrum of a $T_{eff} = 6000\,\text{K}$, $\log_{10}(g\,/\,\text{cms}^{-2}) = 4.0$, [Fe/H] $= -1.0$ star. Its resolution is 1.0 Å (full-width at half-maximum). Line data are from the Moore, Minnaert and Houtgast (1966) list.

Figure 7a. The metallicity sensitivity of the 4900–5000 Å region of the spectrum of a metal-poor G0V star. The graph shows the difference in residual intensity produced by a decrease in metallicity from [Fe/H] $= -1.0$ to $-1.5$ in a $T_{eff} = 6000\,\text{K}$, $\log_{10}(g\,/\,\text{cms}^{-2}) = 4.0$ star as a function of wavelength. The resolution of the spectrum is 1.0 Å (full-width at half-maximum). Line data are from the Moore, Minnaert and Houtgast (1966) list.

Figure 7b. The gravity sensitivity of the 4900–5000 Å region of the spectrum of a metal-poor G0 IV star. The graph shows the difference in residual intensity produced by a change in surface gravity from $\log_{10}(g\,/\,\text{cms}^{-2}) = 3.0$ to 4.0 in a $T_{eff} = 6000\,\text{K}$, [Fe/H] $= 0.0$ star as a function of wavelength. The resolution of the spectrum is 1.0 Å (full-width at half-maximum). Line data are from the Moore, Minnaert and Houtgast (1966) list.

Figure 8a. The metallicity sensitivity of the 4025–4090 Å region of the spectrum of a mildly metal-poor G5V star. The graph shows the difference in residual intensity produced by a decrease in metallicity from [Fe/H] $= 0.0$ to $-0.5$ in a $T_{eff} = 5500\,\text{K}$, $\log_{10}(g\,/\,\text{cms}^{-2}) = 4.0$ star as a function of wavelength. The resolution of the spectrum is 1.0 Å (full-width at half-maximum). Line data are from the Moore, Minnaert and Houtgast (1966) list.

Figure 8b. The gravity sensitivity of the 4025–4090 Å region of the spectrum of a moderately metal-poor G5 IV star. The graph shows the difference in residual intensity produced by a change in surface gravity from $\log_{10}(g\,/\,\text{cms}^{-2}) = 3.0$ to 4.0 in a $T_{eff} = 5500\,\text{K}$, [Fe/H] $= -0.5$ star as a function of wavelength. The resolution of the spectrum is 1.0 Å (full-width at half-maximum). Line data are from the Moore, Minnaert and Houtgast (1966) list.

Figure 9a. The metallicity sensitivity of the 4500–4570 Å region of the spectrum of a mildly metal-poor G5V star. See Figure 8a for details.

Figure 9b. The gravity sensitivity of the 4500–4570 Å region of the spectrum of a moderately metal-poor G5 IV star. See Figure 8b for details.

Figure 10a. The metallicity sensitivity of the 4565–4630 Å region of the spectrum of a mildly metal-poor G5 V star. See Figure 8a for details.

Figure 10b. The gravity sensitivity of the 4565–4630 Å region of the spectrum of a moderately metal-poor G5 IV star. See Figure 8b for details.

Figure 11a. The metallicity sensitivity of the 4625–4690 Å region of the spectrum of a mildly metal-poor G5 V star. See Figure 8a for details.

Figure 11b. The gravity sensitivity of the 4625–4690 Å region of the spectrum of a moderately metal-poor G5 IV star. See Figure 8b for details.

Figure 12a. The metallicity sensitivity of the 4875–4940 Å region of the spectrum of a mildly metal-poor G5 V star. See Figure 8a for details.

Figure 12b. The gravity sensitivity of the 4875–4940 Å region of the spectrum of a moderately metal-poor G5 IV star. See Figure 8b for details.

Figure 13a. The metallicity sensitivity of the 4935–5000 Å region of the spectrum of a mildly metal-poor G5 V star. See Figure 8a for details.

Figure 13b. The gravity sensitivity of the 4935–5000 Å region of the spectrum of a moderately metal-poor G5 IV star. See Figure 8b for details.

Figure 14. The derivation of the [Fe/H] – log $g$ relation corresponding to a particular observed index value. The diagram illustrates the [Fe/H] – log $g$ relation of an index which is defined by a knowledge of the index and the stellar temperature. The temperature $T_{eff}$ defines a two-dimensional surface in the abundance – surface gravity – index parameter space, while the observed index value defines a plane. The two surfaces intersect at a line which defines the abundance – gravity relation of the index.

Figure 15. The [Fe/H] – log $g$ relations of the 13 indices for the noise-free first test spec-

trum. The iron abundance – surface gravity relations derived from the 13 indices of the 4025–4930 Å region are shown for the $T_{eff} = 5200\,\text{K}$, $\log_{10}(g\,/\,\text{cms}^{-2}) = 4.2$, $[\text{Fe/H}] = -0.3$, $\xi_{micro} = 1.5\,\text{kms}^{-1}$ synthetic test spectrum. Relations marked MO are derived from the abundance indices sensitive to metallicity but comparatively insensitive to gravity, those marked AM are derived from the indices showing appreciable sensitivity to both metallicity and gravity, while those labelled IG are obtained from the ionic gravity indices.

Figure 16. The [Fe/H] – log $g$ relations of the 13 indices for the noise-added first test spectrum. The iron abundance – surface gravity relations derived from the 13 indices of the 4025–4930 Å region are shown for the $T_{eff} = 5200\,\text{K}$, $\log_{10}(g\,/\,\text{cms}^{-2}) = 4.2$, $[\text{Fe/H}] = -0.3$, $\xi_{micro} = 1.5\,\text{kms}^{-1}$ test spectrum. Random errors had been added to the intensity data corresponding to a signal-to-noise ratio of 12 in each 1.0 Å wide resolution element. It becomes difficult to estimate a best intersection region.

Figure 17. The [Fe/H] – log $g$ relations of the 13 indices for the noise-added second test spectrum. The iron abundance – surface gravity relations derived from the 13 indices are shown for the $T_{eff} = 5850\,\text{K}$, $\log_{10}(g\,/\,\text{cms}^{-2}) = 3.2$, $[\text{Fe/H}] = -1.2$, $\xi_{micro} = 1.5\,\text{kms}^{-1}$ test spectrum. Random errors had been added to the intensity data corresponding to a signal-to-noise ratio of 12 in each 1.0 Å wide resolution element. It becomes impossible to estimate a best intersection region.

Figure 18. The [Fe/H] – log $g$ relations of the six compound indicators for the noise-free first test spectrum. The iron abundance – surface gravity relations derived from the six compound indicators of the 4025–4930 Å region are shown for the $T_{eff} = 5200\,\text{K}$, $\log_{10}(g\,/\,\text{cms}^{-2}) = 4.2$, $[\text{Fe/H}] = -0.3$, $\xi_{micro} = 1.5\,\text{kms}^{-1}$ test spectrum.

Figure 19. The [Fe/H] – log $g$ relations of the six compound indicators for the noise-added first test spectrum. The iron abundance – surface gravity relations derived from the six compound indicators of the 4025–4930 Å region are shown for the $T_{eff} = 5200\,\text{K}$, $\log_{10}(g\,/\,\text{cms}^{-2}) = 4.2$, $[\text{Fe/H}] = -0.3$, $\xi_{micro} = 1.5\,\text{kms}^{-1}$ test spectrum. Random errors have been added to the intensity data corresponding to a signal-to-noise ratio of 12 in each 1.0 Å wide resolution element. It becomes easier to estimate a best intersection region than from the 13 index analysis results of Figure 16.

Figure 20. The [Fe/H] – log $g$ relations of the six compound indicators for the noise-added second test spectrum. The iron abundance – surface gravity relations derived from the six compound indicators are shown for the $T_{eff} = 5850\,\mathrm{K}$, $\log_{10}(g/\mathrm{cms}^{-2}) = 3.2$, [Fe/H] = $-1.2$, $\xi_{micro} = 1.5\,\mathrm{kms}^{-1}$ test spectrum. Random errors had been added to the intensity data corresponding to a signal-to-noise ratio of 12 in each 1.0 Å wide resolution element. As in the case of Figure 17, it is still impossible to estimate a best intersection region.

Table 2 : The numbers of atomic and ionic lines used to compute synthetic spectra

| Wavelength region ( in Å ) | Number of spectral lines used |
|---|---|
| 4025 – 4092 | 483 |
| 4500 – 4690 | 1215 |
| 4840 – 4870 | 162 |
| 4870 – 5000 | 572 |

The numbers of atomic and ionic spectral lines used in the computation of sections of stellar spectra are listed above, a total of 2432 lines. The $4840 - 4870$Å region was not used to define index flux bands.

Table 3 : The stellar parameters of the basic grid of stellar spectra.

| | | | |
|---|---|---|---|
| Effective temperatures : | $T_{eff}$ | = | 4500, 5000, 5500, 6000, 6500 K |
| Surface gravities : | $\log_{10}(g/\mathrm{cms}^{-2})$ | = | 2.0, 3.0, 4.0, 5.0 |
| Metallicities : | [Fe/H] | = | 0.0, −0.5, −1.0, −1.5, −2.0 |
| Microturbulence : | $\xi_{micro}$ | = | 1.5 kms$^{-1}$ |

The stellar parameters used to generate a basic grid of synthetic spectra are listed above. A total of 100 different stellar models were used. Additional spectra were computed, including some with a microturbulence of $\xi_{micro}$ = 1.0 kms$^{-1}$ for assessing the effect of microturbulence errors.

Table 4 : Molecular contamination in the solar spectrum.

| Molecule | Wavelength region (in Å) | Degree of contamination | Notes |
|---|---|---|---|
| CH | 4000 – 4150 | Slight | Contamination not great, indices still possible. |
|  | 4150 – 4420 | Moderate to very strong | Indices not possible |
|  | 4870 – 4890 | Very slight | Insignificant |
| CN | 4080 – 4130 | Slight | Insignificant |
|  | 4130 – 4170 | Moderate to strong | Indices not possible |
|  | 4170 – 4200 | Very strong | Indices not possible |
|  | 4200 – 4220 | Strong | Indices not possible |
| $C_2$ | 4650 - 4680 | Slight | Insignificant |
|  | 4680 – 4740 | Slight | Comparatively unimportant |
|  | 4830 – 5000 | Very slight | Very few lines — insignificant |
| MgH | 4950 – 5000 | Slight | Insignificant |

Table 5 : A provisional list of abundance indicators.

| Index number | Absorption band wavelength range ( in Å) | Comparison band wavelength range ( in Å) |
|---|---|---|
| 1 | 4043.5 – 4047.5 | 4037.0 – 4043.5 |
|   |                 | 4047.5 – 4051.0 |
| 2 | 4062.0 – 4065.5 | 4080.5 – 4082.5 |
|   | 4070.0 – 4073.5 | 4087.0 – 4091.5 |
| 3 | 4524.5 – 4537.0 | 4503.0 – 4511.5 |
|   |                 | 4518.5 – 4521.5 |
| 4 | 4578.0 – 4587.5 | 4573.0 – 4577.5 |
|   | 4591.0 – 4606.0 | 4587.5 – 4591.5 |
|   |                 | 4608.0 – 4610.5 |
| 5 | 4678.5 – 4683.0 | 4658.0 – 4663.0 |
| 6 | 4883.2 – 4893.2 | 4893.5 – 4902.0 |
| 7 | 4917.5 – 4925.5 | 4913.0 – 4917.5 |
| 8 | 4935.0 – 4940.0 | 4940.5 – 4945.0 |
|   |                 | 4947.5 – 4952.0 |
| 9 | 4955.0 – 4959.5 | 4959.5 – 4964.5 |
| 10 | 4977.5 – 4987.0 | 4987.0 – 4990.0 |
|    |                 | 4973.5 – 4977.5 |

Table 8 : The iron abundance indices selected.

| Index number | Absorption band wavelength limits (in Å) | Comparison band wavelength limits (in Å) | Sensitivity of index to gravity |
|---|---|---|---|
| 1. | 4977.5 – 4986.6 | 4960.0 – 4964.5<br>4971.4 – 4975.4<br>4986.6 – 4992.5 | None |
| 2. | 4935.9 – 4940.3 | 4949.0 – 4952.0<br>4940.3 – 4945.1<br>4925.8 – 4931.6 | None |
| 3. | 4885.1 – 4889.4<br>4909.2 – 4912.3 | 4877.4 – 4881.1<br>4895.0 – 4899.5<br>4901.3 – 4909.2<br>4912.3 – 4917.6 | None |
| 4. | 4956.4 – 4958.9 | 4949.0 – 4952.0<br>4960.0 – 4964.5 | Strong |
| 5. | 4917.6 – 4922.5 | 4901.3 – 4909.2<br>4912.3 – 4917.6<br>4925.8 – 4931.6 | Strong |
| 6. | 4870.0 – 4873.0<br>4890.2 – 4892.7 | 4877.4 – 4881.8<br>4895.0 – 4899.5 | Strong |
| 7. | 4665.5 – 4669.0<br>4672.0 – 4675.4<br>4677.9 – 4682.8 | 4650.5 – 4654.1<br>4658.2 – 4662.1<br>4675.4 – 4677.9 | None |
| 8. | 4636.9 – 4641.4 | 4623.5 – 4627.6<br>4630.0 – 4633.7<br>4642.9 – 4645.8 | None |
| 9. | 4590.9 – 4601.4<br>4578.7 – 4582.0<br>4563.8 – 4567.2 | 4556.4 – 4563.8<br>4572.6 – 4578.4<br>4587.3 – 4590.9 | None |
| 10. | 4512.0 – 4519.1<br>4522.3 – 4530.4<br>4538.5 – 4542.9 | 4502.8 – 4511.5<br>4519.1 – 4522.3<br>4556.4 – 4563.8 | None |
| 11. | 4043.5 – 4047.9<br>4052.0 – 4058.6<br>4061.8 – 4065.2<br>4070.1 – 4073.3 | 4036.6 – 4043.5<br>4080.5 – 4089.9 | Strong |

The iron abundance indices are listed in the table above. They have been selected from the flux bands of Table 6.

Table 9 : The ionic gravity indices selected.

| Index number | Ionic absorption band wavelength limits (in Å) | Comparison band wavelength limits (in Å) |
|---|---|---|
| 1. | 4873.0 – 4878.0 | 4885.1 – 4889.4 |
|    | 4892.0 – 4895.0 | 4895.0 – 4899.5 |
|    | 4909.7 – 4912.7 | 4901.3 – 4909.2 |
|    | 4922.5 – 4925.0 | |
| 2. | 4500.0 – 4503.0 | 4503.0 – 4506.0 |
|    | 4506.0 – 4509.0 | 4509.0 – 4514.0 |
|    | 4514.0 – 4516.5 | 4516.5 – 4519.1 |
|    | 4519.1 – 4525.0 | 4525.5 – 4528.0 |
|    | 4531.0 – 4534.5 | |
|    | 4540.5 – 4546.0 | |
|    | 4548.0 – 4559.5 | |
| 3. | 4562.0 – 4565.0 | 4578.7 – 4582.0 |
|    | 4567.0 – 4569.5 | 4585.0 – 4587.5 |
|    | 4571.0 – 4573.5 | 4596.5 – 4599.5 |
|    | 4575.0 – 4578.0 | 4602.0 – 4605.5 |
|    | 4582.0 – 4585.0 | |
|    | 4587.5 – 4593.0 | |
| 4. | 4608.0 – 4610.5 | 4602.0 – 4605.5 |
|    | 4615.5 – 4621.5 | 4621.5 – 4624.0 |
|    | 4628.0 – 4630.5 | 4630.0 – 4633.7 |
|    | 4633.5 – 4636.9 | 4636.9 – 4641.4 |
|    | | 4642.9 – 4645.8 |
| 5. | 4647.5 – 4650.0 | 4643.5 – 4647.5 |
|    | 4655.9 – 4658.4 | 4652.0 – 4655.9 |
|    | 4661.5 – 4667.5 | 4672.0 – 4677.8 |

The ionic gravity indices are listed above. They have been selected from the flux bands of Table 7.

Table 10 : The six compound indicators for cursor method analyses.

| Compound indicator number | Nature of indicator | Indices of Table 8 used to define indicator | Indices of Table 9 used to define indicator |
|---|---|---|---|
| 1 | Metallicity sensitive | 1, 2, 3 | – |
| 2 | Metallicity sensitive | 7, 8 | – |
| 3 | Metallicity sensitive | 9, 10 | – |
| 4 | Metallicity and gravity sensitive | 4, 5, 6, 11 | – |
| 5 | Ionic gravity | – | 1, 4, 5 |
| 6 | Ionic gravity | – | 2, 3 |

Table 11 : The noise-induced random error in the [Fe/H] results of the comprehensive compound indicator sensitive to both abundance and gravity.

| [Fe/H] | $T_{eff}$ (in K) | Error in [Fe/H] due to noise in spectrum for S/N ratios (1.0 Å pixels) of ||||||
|---|---|---|---|---|---|---|---|
| | | 5 | 10 | 20 | 30 | 50 | 80 |
| 0.0 | 4500 | 0.40 | 0.17 | 0.075 | 0.046 | 0.025 | 0.014 |
| 0.0 | 5000 | 0.41 | 0.18 | 0.078 | 0.048 | 0.026 | 0.015 |
| 0.0 | 5500 | 0.41 | 0.18 | 0.080 | 0.049 | 0.027 | 0.015 |
| 0.0 | 6000 | 0.42 | 0.19 | 0.082 | 0.051 | 0.028 | 0.016 |
| 0.0 | 6500 | 0.43 | 0.19 | 0.084 | 0.053 | 0.029 | 0.017 |
| -0.5 | 4500 | 0.43 | 0.19 | 0.086 | 0.053 | 0.030 | 0.017 |
| -0.5 | 5000 | 0.46 | 0.21 | 0.098 | 0.062 | 0.035 | 0.021 |
| -0.5 | 5500 | 0.49 | 0.23 | 0.11 | 0.073 | 0.043 | 0.026 |
| -0.5 | 6000 | 0.52 | 0.26 | 0.13 | 0.085 | 0.051 | 0.032 |
| -0.5 | 6500 | 0.55 | 0.29 | 0.15 | 0.10 | 0.061 | 0.039 |
| -1.0 | 4500 | 0.45 | 0.21 | 0.097 | 0.062 | 0.035 | 0.021 |
| -1.0 | 5000 | 0.51 | 0.25 | 0.12 | 0.082 | 0.049 | 0.030 |
| -1.0 | 5500 | 0.57 | 0.30 | 0.16 | 0.11 | 0.067 | 0.043 |
| -1.0 | 6000 | 0.64 | 0.36 | 0.20 | 0.14 | 0.093 | 0.063 |
| -1.0 | 6500 | 0.72 | 0.43 | 0.26 | 0.19 | 0.13 | 0.091 |
| -1.5 | 4500 | 0.48 | 0.23 | 0.11 | 0.072 | 0.042 | 0.025 |
| -1.5 | 5000 | 0.57 | 0.30 | 0.16 | 0.11 | 0.066 | 0.043 |
| -1.5 | 5500 | 0.67 | 0.39 | 0.22 | 0.16 | 0.11 | 0.073 |
| -1.5 | 6000 | 0.79 | 0.50 | 0.31 | 0.24 | 0.17 | 0.12 |
| -1.5 | 6500 | 0.94 | 0.65 | 0.45 | 0.36 | 0.27 | 0.21 |
| -2.0 | 4500 | 0.51 | 0.25 | 0.13 | 0.083 | 0.049 | 0.031 |
| -2.0 | 5000 | 0.64 | 0.35 | 0.20 | 0.14 | 0.091 | 0.061 |
| -2.0 | 5500 | 0.79 | 0.50 | 0.31 | 0.24 | 0.17 | 0.12 |
| -2.0 | 6000 | 0.98 | 0.69 | 0.49 | 0.40 | 0.31 | 0.25 |
| -2.0 | 6500 | 1.22 | 0.97 | 0.77 | 0.68 | 0.57 | 0.49 |

The smoothed estimates of the random error in the [Fe/H] results are presented for various signal-to-noise ratios and for different stellar effective temperatures and metallicities.

Table 12 : The random errors in effective temperature estimates produced by errors in $(V–I)_C$ colour indices.

| | Error in $T_{eff}$ (in K) resulting from errors in $(V-I)_C$ of | | |
|---|---|---|---|
| $T_{eff}$ (in K) | $\pm 0\overset{m}{\cdot}05$ | $\pm 0\overset{m}{\cdot}02$ | $\pm 0\overset{m}{\cdot}01$ |
| 4500 | $\pm 120$ | $\pm 47$ | $\pm 23$ |
| 5000 | 140 | 58 | 29 |
| 5500 | 170 | 70 | 35 |
| 6000 | 210 | 83 | 42 |
| 6500 | 240 | 97 | 49 |

Table 13 : The errors in [Fe/H] results caused by photometric errors.

| Nominal [Fe/H] value | Nominal effective temperature (in K) | Error in the [Fe/H] result produced by an error in $(V-I)_C$ of | | |
|---|---|---|---|---|
| | | $\pm 0^{m}\!.05$ | $\pm 0^{m}\!.02$ | $\pm 0^{m}\!.01$ |
| 0.0 | 5000 | $\pm 0.13$ | $\pm 0.05$ | $\pm 0.03$ |
| | 6000 | 0.14 | 0.06 | 0.03 |
| $-1.0$ | 5000 | 0.14 | 0.06 | 0.03 |
| | 6000 | 0.16 | 0.06 | 0.03 |
| $-2.0$ | 5000 | 0.16 | 0.06 | 0.03 |
| | 6000 | 0.17 | 0.07 | 0.04 |

The errors in the [Fe/H] result which would be produced by different errors in the $(V-I)_C$ colour index are presented for several types of star.

Table 14 : The error in [Fe/H] produced by an error of +1.0 in $\log_{10} g$.

| Nominal [Fe/H] value | Error in the [Fe/H] result caused by an error of +1.0 in $\log_{10} g$ for a temperature of | | | | |
|---|---|---|---|---|---|
| | 4500 K | 5000 K | 5500 K | 6000 K | 6500 K |
| 0.0 | −0.12 | −0.14 | −0.14 | −0.11 | −0.07 |
| −0.5 | −0.14 | −0.16 | −0.15 | −0.10 | −0.06 |
| −1.0 | −0.17 | −0.18 | −0.16 | −0.11 | −0.07 |
| −1.5 | −0.24 | −0.25 | −0.23 | −0.18 | −0.15 |
| −2.0 | −0.31 | −0.26 | −0.20 | −0.14 | −0.12 |

Table 15 : The errors in [Fe/H] results produced by a $0.5\,\text{kms}^{-1}$ error in the microturbulence parameter.

| Nominal [Fe/H] value | Error in the [Fe/H] result caused by an error of $+0.5\,\text{kms}^{-1}$ in $\xi_{micro}$ at $\xi_{micro} = 1.25\,\text{kms}^{-1}$ for a temperature of | | |
|---|---|---|---|
| | 5000 K | 5500 K | 6000 K |
| 0.0 | + 0.17 | 0.12 | 0.08 |
| −1.5 | + 0.04 | 0.03 | 0.05 |

Table 16 : The errors in [Fe/H] produced by errors in the wavelength scale calibrations for different metallicities.

| Wavelength scale error (in Å) | Mean value of the error in [Fe/H] for temperatures of | | |
|---|---|---|---|
| | 5000 K | 5500 K | 6000 K |

For [Fe/H] = 0.0 :

| | | | |
|---|---|---|---|
| 0.05 | 0.00 | 0.00 | 0.00 |
| 0.10 | −0.01 | 0.00 | −0.01 |
| 0.20 | −0.02 | −0.02 | −0.02 |
| 0.4 | −0.10 | −0.08 | −0.06 |
| 0.6 | −0.21 | −0.15 | −0.12 |
| 0.8 | −0.34 | −0.25 | −0.21 |
| 1.0 | −0.48 | −0.37 | −0.30 |

For [Fe/H] = −1.0 :

| | | | |
|---|---|---|---|
| 0.05 | 0.00 | 0.00 | 0.00 |
| 0.10 | 0.00 | 0.00 | 0.00 |
| 0.20 | −0.01 | −0.01 | −0.01 |
| 0.4 | −0.06 | −0.04 | −0.04 |
| 0.6 | −0.10 | −0.08 | −0.06 |
| 0.8 | −0.18 | −0.14 | −0.12 |
| 1.0 | −0.27 | −0.21 | −0.19 |

For [Fe/H] = −2.0 :

| | | | |
|---|---|---|---|
| 0.05 | 0.00 | 0.00 | 0.00 |
| 0.10 | 0.00 | 0.00 | 0.00 |
| 0.20 | −0.02 | 0.00 | 0.00 |
| 0.4 | −0.04 | −0.04 | −0.05 |
| 0.6 | −0.08 | −0.06 | −0.07 |
| 0.8 | −0.12 | −0.14 | −0.09 |
| 1.0 | −0.17 | −0.16 | −0.17 |

Table 1 : Adopted collisional broadening enhancement factors.

| Species | Adopted enhancement | References |
| --- | --- | --- |
| Na I | 1.5 | Gratton & Sneden, 1988, Bell *et al.*, 1985, Edvardsson, 1983. |
| Mg I | 2.1 | Gratton & Sneden, 1988. |
| Al I | 1.3 | Gratton & Sneden, 1988. |
| Si I | 1.9 | Gratton & Sneden, 1988, Magain, 1985. |
| Ca I | 1.5 | Gratton & Sneden, 1988, Magain, 1985, O'Neil & Smith, 1980, S Bell *et al.*, 1985. |
| Ca II | 1.2 | Magain, 1985, Gratton & Sneden, 1988, Holweger, 1972. |
| Ti I | 1.08 | Blackwell *et al.*, 1987. |
| Cr I | 1.23 | Blackwell *et al.*, 1987. |
| Fe I | 1.0 to 1.4 | Discussed in Section 2.3. |
| Ni I | 1.75 | Magain, 1985. |
| Sr II | 1.0 | Gratton & Sneden, 1988. |

The enhancement factors chosen for each species for the computation of synthetic spectra are presented. Where the literature. Fe I data are discussed in detail elsewhere. The Edvardsson (1983) result is quoted by Magain ( (1972) is quoted by Gratton & Sneden (1988). Default enhancements of $E = 1.5$ were adopted for lines of specie enhancement data were available. These were C I, S I, K I, Sc I, Sc II, Ti II, V I, V II, Cr II, Mn I, Mn II, Fe II, Sr I, Y I, Y II, Zr I, Zr II, Nb I, Mo I, Ru I, In I, Ba II, La II, Ce II, Pr II, Nd II, Sm II, Eu I, Gd II, Dy II, Hf default value was also adopted for Ca I because available published data show a large scatter. A total of 50 absor

Table 6 : The flux bands chosen to form iron abundance indices.

| Wavelength limits of band (in Å) | Nature of flux band | Sensitivity to gravity | Comments on the nature of the band in |
|---|---|---|---|
| 4036.3 – 4043.5 | Comparison | Slight − | Region of moderate abs. Some Fe I abs. Sligh |
| 4043.5 – 4047.9 | Absorption | Strong + | One v. strong Fe I line (4045.8Å). Some weak |
| 4052.0 – 4058.6 | Absorption | Slight − | Several strong Fe I lines. Other species presen |
| 4061.8 – 4065.2 | Absorption | Strong + | One v. strong Fe I line (4063.6Å). |
| 4070.1 – 4073.3 | Absorption | Strong + | One v. strong Fe I line (4071.7Å). |
| 4080.5 – 4089.9 | Comparison | Slight − | Region of moderate abs. |
| 4502.8 – 4511.5 | Comparison | Slight − | Region of low absorption. |
| 4512.0 – 4519.1 | Absorption | Slight − | Abs. from lines on linear part of curve of grov |
| 4519.1 – 4522.3 | Comparison | − | Low abs. region. Fe II line present. |
| 4522.3 – 4530.4 | Absorption | Some − | Several strong Fe I lines, also other species. |
| 4538.5 – 4542.9 | Absorption | Slight − | Fe & Cr rich. |
| 4556.4 – 4563.8 | Comparison | Some − | Low abs. region. |
| 4563.8 – 4567.2 | Absorption | − | Fe I and other species present. Some weak ior |
| 4572.6 – 4578.4 | Comparison | − | Region of comparatively low absorption. |
| 4578.7 – 4582.0 | Absorption | None | Several Fe I lines. |
| 4587.3 – 4590.9 | Comparison | Slight − | Region of fairly low absorption. |
| 4590.9 – 4601.4 | Absorption | Slight − | Many Fe I lines, some Cr I & Ni I also. |
| 4623.5 – 4627.6 | Comparison | Slight − | Region of comparatively little absorption. |
| 4630.0 – 4633.7 | Comparison | None | Region of comparatively little absorption. |
| 4636.9 – 4641.4 | Absorption | None | Region of strong Fe I abs., some Ti I. |
| 4642.9 – 4645.8 | Comparison | None | Low absorption region. |
| 4650.5 – 4654.1 | Comparison | None | Low absorption region. |
| 4658.2 – 4662.1 | Comparison | Slight − | Region of fairly low absorption. |
| 4665.5 – 4669.0 | Absorption | − | Several lines of Fe I, some Cr I. |

Table 6 continued.

| Wavelength limits of band ( in Å) | Nature of flux band | Sensitivity to gravity | Comments on the nature of the band in |
|---|---|---|---|
| 4672.0 – 4675.4 | Absorption | None | Several Fe I lines. |
| 4675.4 – 4677.9 | Comparison | None | Low absorption region. |
| 4677.9 – 4682.8 | Absorption | Little − | Several Fe I lines. |
| 4870.0 – 4873.0 | Absorption | + | Two very strong Fe I lines. Close to H$\beta$ line |
| 4877.4 – 4881.1 | Comparison | None | Fairly empty region. |
| 4885.1 – 4889.4 | Absorption | None | Strong Fe I, some Cr I. |
| 4890.2 – 4892.7 | Absorption | + | Two very strong Fe I lines. |
| 4895.0 – 4899.5 | Comparison | None | Low absorption region. |
| 4901.3 – 4909.2 | Comparison | None | Fairly empty region. |
| 4909.2 – 4912.3 | Absorption | − | Several strong Fe I lines. |
| 4912.3 – 4917.6 | Comparison | None | Region of comparatively little absorption. |
| 4917.6 – 4922.5 | Absorption | + | Two very strong Fe I lines. |
| 4925.8 – 4931.6 | Comparison | None | Some Fe I abs. present. |
| 4935.9 – 4940.3 | Absorption | Slight + | Several strong Fe I lines. |
| 4940.3 – 4945.1 | Comparison | None | Low absorption region. |
| 4949.0 – 4952.0 | Comparison | None | Low absorption region. |
| 4956.4 – 4958.9 | Absorption | + | Two very strong Fe I lines. |
| 4960.0 – 4964.5 | Comparison | Slight − | Region of relatively low abs. |
| 4971.4 – 4975.4 | Comparison | None | Region of relatively low abs. |
| 4977.5 – 4986.6 | Absorption | None | Strong Fe I abs. |
| 4986.6 – 4992.5 | Comparison | None | Region of fairly low abs. |

Table 6 concluded.

The flux bands chosen to serve as the bases for indices sensitive to metallicity (primarily iron abundance) are lis
in turn the wavelength limits, the nature of the band (whether absorption or comparison), the sensitivity of th
comments relating to the contents of the band in the solar spectrum.

The abbreviations used are :

- abs. - absorption
- v. - very
- + - absorption in band increases with gravity
- − - absorption in band decreases with gravity

Table 7 : The flux bands chosen to form ionic gravity indices.

| Wavelength limits of band (in Å) | Nature of flux band | Comments on the nature of the band in solar spectrum |
|---|---|---|
| 4500.0 – 4503.0 | Gravity | Strong Ti II line |
| 4503.0 – 4506.0 | Comparison | Region of low absorption |
| 4506.0 – 4509.0 | Gravity | Strong Fe II, weak Ti II |
| 4509.0 – 4514.0 | Comparison | Region of fairly low absorption |
| 4514.0 – 4516.5 | Gravity | Strong Fe II |
| 4516.5 – 4519.1 | Comparison | Region free from ionic lines |
| 4519.1 – 4525.0 | Gravity | Fe II line, some Ti II, some minor species |
| 4525.5 – 4528.0 | Comparison | Region of fairly strong abs., but no ionic lines, little gravity sens. |
| 4531.0 – 4534.5 | Gravity | Strong Ti II and Fe II lines. |
| 4540.5 – 4546.0 | Gravity | Fe II and Ti II lines |
| 4548.0 – 4559.5 | Gravity | Strong Fe II abs., also Ti II & Cr II (and a Ba I |
| 4562.0 – 4565.0 | Gravity | Strong Ti II |
| 4567.0 – 4569.5 | Gravity | Ti II line |
| 4571.0 – 4573.5 | Gravity | Strong Ti II line |
| 4575.0 – 4578.0 | Gravity | Fe II line |
| 4578.7 – 4582.0 | Comparison | Region of strong Fe I abs., but little grav. sens. |
| 4582.0 – 4585.0 | Gravity | Strong Fe II abs., some Ti II |
| 4585.0 – 4587.5 | Comparison | Region of low grav. sens., contaminated by Ca I |
| 4587.5 – 4593.0 | Gravity | Cr II & Ti II abs. |
| 4596.5 – 4599.5 | Comparison | Some Fe I abs. |
| 4602.0 – 4605.5 | Comparison | Region of fairly low abs., free from ionic lines |
| 4608.0 – 4610.5 | Gravity | Contains a weak Ti II line. |
| 4615.5 – 4621.5 | Gravity | Cr II and Fe II lines |

Table 7 continued.

| Wavelength limits of band (in Å) | Nature of flux band | Comments on the nature of the band solar spectrum |
|---|---|---|
| 4621.5 – 4624.0 | Comparison | Some Cr I abs |
| 4628.0 – 4630.5 | Gravity | Fe II & Ce II abs. |
| 4630.5 – 4633.7 | Comparison | Low abs. region |
| 4633.5 – 4636.9 | Gravity | Cr II, Fe II & Ti II lines |
| 4636.9 – 4641.4 | Comparison | Strong Fe I & Ti I abs. |
| 4642.9 – 4645.8 | Comparison | Low abs. region no ionic lines |
| 4643.5 – 4647.5 | Comparison | Strong Cr I abs., little ionic abs. |
| 4647.5 – 4650.0 | Gravity | Fe II line (some Cr I abs.) |
| 4652.0 – 4655.9 | Comparison | Comparatively low abs. region |
| 4655.9 – 4658.4 | Gravity | Fe II & Ti II abs. |
| 4661.5 – 4667.5 | Gravity | Fe II & Ti II abs. |
| 4672.0 – 4677.8 | Comparison | Some Fe I abs., some contamination by $C_2$ and unidentified lines |
| 4873.0 – 4878.0 | Gravity | Ti II & Cr II abs. |
| 4885.1 – 4889.4 | Comparison | Considerable Fe I abs., little gravity sensitivity |
| 4892.0 – 4895.0 | Gravity | Weak Fe II abs. |
| 4895.0 – 4899.5 | Comparison | Very little abs. |
| 4901.3 – 4909.2 | Comparison | Little abs., insensitive to gravity |
| 4909.7 – 4912.7 | Gravity | Strong Fe I abs., some Ti II |
| 4922.5 – 4925.0 | Gravity | Strong Fe II line |

Table 7 concluded.

The flux bands chosen to serve as the bases for ionic gravity indices are listed above. The columns give in turn the of the band (whether gravity sensitive or comparison), and comments relating to the contents of the band in the

The abbreviations used are :

| | | |
|---|---|---|
| abs. | - | absorption |
| v. | - | very |
| sens. | - | sensitivity |
| grav. | - | gravity |